\documentclass[12pt,preprint,useAMS,usenatbib]{emulateapj}
\usepackage{graphicx}
\def\OVI{\hbox{O~$\scriptstyle\rm VI$}}
\def\msun{\,{\rm M_\odot}}

\def\rvir{R_{\rm vir}\,}
\def\sfr{\,{\rm M_\odot\,yr^{-1}}}

\def\spose#1{\hbox to 0pt{#1\hss}}
\def\lta{\mathrel{\spose{\lower 3pt\hbox{$\mathchar"218$}}
     \raise 2.0pt\hbox{$\mathchar"13C$}}}
\def\gta{\mathrel{\spose{\lower 3pt\hbox{$\mathchar"218$}}
     \raise 2.0pt\hbox{$\mathchar"13E$}}}

\newcommand{\etal}{{et al.\ }}

\def\CIV{\hbox{C~$\scriptstyle\rm IV$}}

\def\CII{\hbox{C~$\scriptstyle\rm II$}}

\def\OVI{\hbox{O~$\scriptstyle\rm VI$}}

\def\SiII{\hbox{Si~$\scriptstyle\rm II$}}

\def\SiIV{\hbox{Si~$\scriptstyle\rm IV$}}
\def\kms{\,{\rm km\,s^{-1}}}

\def\cmm{\,{\rm cm^{-2}}}

\def\sfr{\,{\rm M_\odot\,yr^{-1}}}

\def\Lya{Ly$\alpha$}

\lefthead{Shen et al. 2011}
\righthead{Metals in circumgalactic medium}
\submitted{submitted to the ApJ}


\begin{document}

\title{The origin of metals in the circumgalactic medium of massive galaxies at $z=3$}
\author{Sijing~Shen$^1$, Piero~Madau$^{1,2}$, Anthony~Aguirre $^{1}$, Javiera~Guedes$^{1,3}$, Lucio Mayer$^4$, and James~Wadsley$^{5}$}

\altaffiltext{1}{Department of Astronomy and Astrophysics, University of California, Santa Cruz, 1156 High Street, Santa Cruz, CA 95064.}
\altaffiltext{2}{Institute of Astronomy, Madingley Road, Cambridge CB3 0HA, United Kingdom.}
\altaffiltext{3}{Institute for Astronomy, ETH Zurich, Wolgang-Pauli-Strasse 27, 8093 Zurich, Switzerland.}
\altaffiltext{4}{Institute of Theoretical Physics, University of Zurich, Winterthurerstrasse 190, CH-9057 Zurich, Switzerland.}
\altaffiltext{5}{Department of Physics and Astronomy, McMaster University, Main Street West, Hamilton L8S 4M1, Canada.}

\begin{abstract}
We present a detailed study of the metal-enriched circumgalactic medium (CGM) of a massive galaxy at $z=3$ using results from ``ErisMC",
a  new cosmological hydrodynamic ``zoom-in" simulation of a disk galaxy with mass comparable to the Milky Way.
The reference run adopts a blastwave scheme for supernova feedback that generates galactic outflows without explicit wind particles, a 
star formation recipe based on a high gas density threshold, and high temperature metal cooling.  
ErisMC main progenitor at $z=3$ resembles a ``Lyman break" galaxy of total mass $M_{\rm vir}=2.4\times 10^{11}\,\msun$, virial radius $\rvir=48$ kpc, and 
star formation rate $18\,\sfr$, and its metal-enriched CGM extends as far as 200 (physical) kpc from its center.
Approximately 41\%, 9\%, and 50\% of all gas-phase metals at $z=3$ are locked in a hot ($T>3\times 10^5$ K), warm ($3\times 10^5 {\rm K}>T>3\times 
10^4$ K), and cold ($T<3\times 10^4$ K) medium, respectively. We identify three sources of 
heavy elements: 1) the main host, responsible for 60\% of all the metals found within $3\rvir$; 2) its satellite progenitors, 
which shed their metals before and during infall, and are responsible for 28\% of all the metals within $3\rvir$, and for only 5\% of 
those beyond $3\rvir$; and nearby dwarfs, which give origin to 12\% of all the metals within $3\rvir$ and 95\% of 
those beyond $3\rvir$. Late ($z<5$) galactic ``superwinds" -- the result of recent star formation in ErisMC -- account for 
only 9\% of all the metals observed beyond $2\rvir$, the bulk having been released at redshifts $5\lta z \lta 8$ by early star formation and outflows. 
In the CGM, lower overdensities are typically enriched by `older', colder metals. Heavy elements are accreted onto ErisMC 
along filaments via low-metallicity cold inflows, and are ejected hot via galactic outflows at a few hundred $\kms$. The outflow mass-loading factor is of
order unity for the main halo, but can exceed a value of 10 for nearby dwarfs. We stress that our ``zoom-in" simulation focuses on the CGM of a 
single massive system and cannot describe the enrichment history of the intergalactic medium as a whole by a population of galaxies with different 
masses and star formation histories. 
%and shows no correlation with the mass of the host over the interval $5.0\times 10^9\,\msun \lta M_{\rm vir} \lta 2.4\times 10^{11}\,\msun$. 
\end{abstract}

\keywords{galaxies: evolution -- galaxies: high-redshift -- intergalactic medium -- method: numerical}

\section{Introduction} \label{intro}

Studies of the ionization, thermodynamic, and kinematic state of heavy elements in circumgalactic gas hold 
clues to understanding the exchange of mass, metals, and energy between galaxies and their surroundings. The 
distribution of observed metals in the intergalactic medium (IGM) is highly inhomogeneous, with a global cosmic abundance at $z=3$ of 
[C/H]$=-2.8\pm 0.13$ for gas with overdensities $0.3<\delta<100$ \citep{Schaye03}. The characteristic epoch of this 
enrichment and its main donors remain uncertain. Late supernova-driven ``superwinds"  from massive galaxies 
\citep[e.g.,][]{Aguirre01,Adelberger03}, early outflows from dwarf galaxies \citep[e.g.,][]{Dekel86,MacLow99,Madau01,Mori02,Scannapieco02,Furlanetto03}, 
quasar-driven winds \citep[e.g.,][]{Scannapieco04}, and the ejection of gas during the merging of protogalaxies \citep{Gnedin98} are all likely to have 
left some chemical imprint on circumgalactic and intergalactic gas, but their relative contributions are not constrained by present data, and 
involve complex baryonic processes that are difficult to model.

Observations clearly show that galactic-scale outflows with velocities of several hundred $\kms$ are ubiquitous in massive star-forming galaxies
at high redshift and in starburst galaxies in the local universe \citep[e.g.,][]{Heckman90,Pettini01,Martin05,Veilleux05,Weiner09,Steidel10}. 
They also indicate that the metal content of the $z\sim 3$ IGM is closely correlated with the positions of nearby galaxies: \OVI\ systems 
are found to be strongly associated with known Lyman-break galaxies (LBGs), most detectable \CIV\ systems ($N_{\rm CIV}\gta 10^{11.5}\,\cmm$) lie 
within 1 proper Mpc from an LBG, and roughly one-third of all ``intergalactic" absorption lines with $N_{\rm CIV}\gta 10^{14}\,\cmm$ are 
produced by gas that lies within $\sim $80 proper kpc from an LBGs \citep{Adelberger03,Adelberger05a}. Somewhat puzzling, the comoving mass density of \CIV\ 
ions in the IGM is found to drop by a factor of 4 at $z\gta 5$ \citep{Ryan-Weber09,Simcoe11}. 

Understanding the galaxy-IGM ecosystem, i.e. the role of inflows, outflows, star formation and active galactic nucleus (AGN) ``feedback" 
in governing the gaseous and metal content of galaxies and their environment, is the goal of many recent theoretical efforts. Hydrodynamical 
simulations of galaxy formation over large cosmological volumes, while advancing rapidly over the past few years 
\citep[e.g.][]{Cen05,Oppenheimer08,Oppenheimer09, Wiersma09,Shen10,Cen11}, typically suffer from poor spatial and 
mass resolution -- which limits their ability to follow self-consistently the 
venting of metals by small galaxies and the transport of heavy elements from their production sites into the environment.    
And while with suitable wind prescriptions these simulations have shown that galactic outflows can potentially enrich the IGM to approximately the observed 
levels, it is not clear whether the same prescriptions are also able to produce realistically looking galaxies. To shed some light on the objects and processes 
responsible for seeding the circumgalactic and intergalactic medium with nuclear waste,
we follow here a different approach. We present results from a new cosmological $N$-body/smooth particle hydrodynamic (SPH) simulation of extreme
dynamic range, a twin of the ``Eris" simulation \citep{Guedes11}. Termed ``ErisMC", the simulations follows the assembly of a
massive galaxy halo with a spline softening length of 120 pc and 26 million dark matter and SPH particles in the high-resolution region. The feedback from an active 
galactic nucleus is neglected, and a star formation recipe is adopted based on a high gas density threshold. 
Heating by supernovae occurs in a clustered fashion, and the resulting pressure-driven outflows at high redshifts 
remove low angular momentum metal-enriched gas. As shown below, our detailed study of ErisMC's CGM at $z=3$ 
reveals that late galactic superwinds -- the result of recent star formation -- account for only a small fraction of 
all the metals found between $\sim$ 100 and 200 (proper) kpc from the center of the main host, and resolves two other sources for the heavy 
elements found in ErisMC's environment: its satellite progenitors -- which deposit their metals before and during infall -- and the orbiting nearby dwarfs.

The paper is organized as follows. In \S~2 we describe the ErisMC simulation. \S~3 presents a detailed study of the galaxy CGM.
The origin of circumgalactic metals and the role played by satellite progenitors and nearby dwarfs in polluting ErisMC's environment are discussed in 
\S~4. The properties of the supernova-driven outflows from the main host are analyzed in \S~5. Finally, in \S~6 we summarize our main results. 
Throughout this work, we use the metal mass fraction $Z_\odot=0.0142$ for the solar abundance \citep{Asplund09}.  

\begin{figure}
\centering
\includegraphics[width=0.48\textwidth]{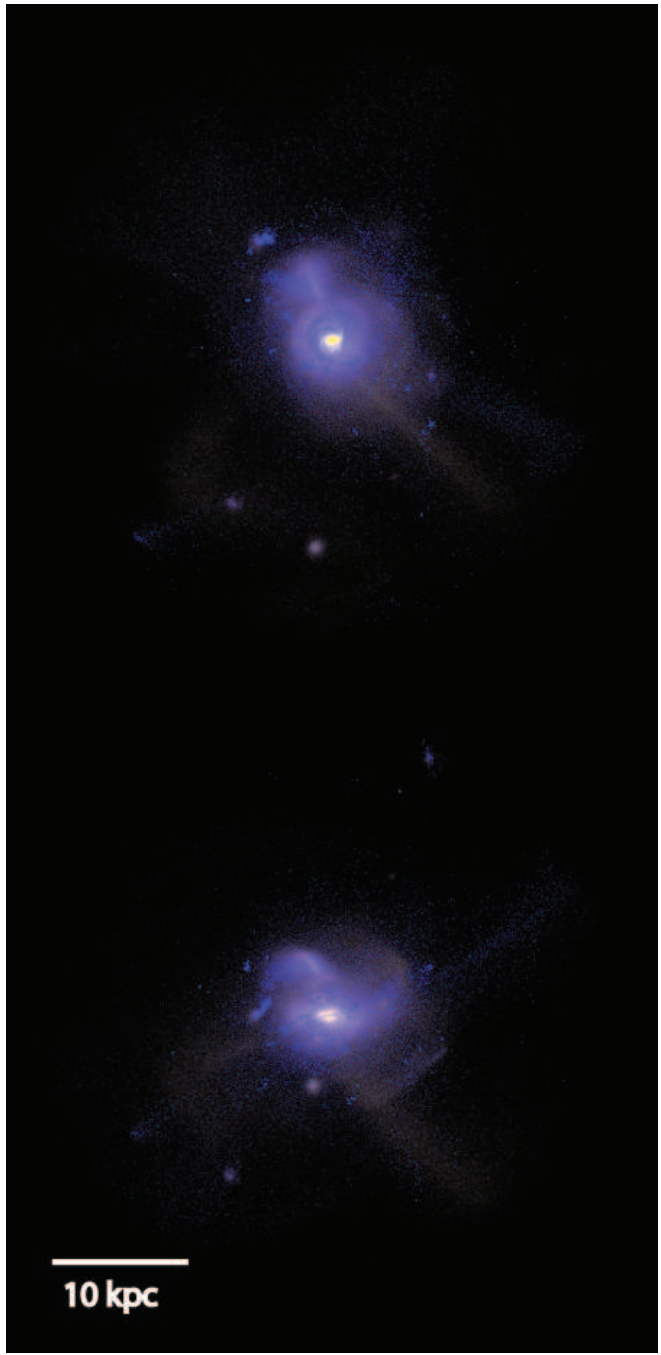}
\vspace{+0.2cm}
\caption{The optical/UV stellar properties of ErisMC at $z=3$. The images were created with the radiative transfer code
{\sc Sunrise} \citep{Jonsson06}, and show a rest-frame $B$, $U$, and $NUV$ stellar composite of the simulated galaxy seen face-on ({\it top panel}) and 
and edge-on ({\it bottom panel}). The 10 kpc bar in the bottom panel is in proper units.
}
\label{fig1}
\vspace{+0.1cm}
\end{figure}

\section{The ErisMC simulation} 
\label{simulation}

The Eris suite of simulations was performed in a {\it Wilkinson Microwave Anisotropy Probe} 3-year cosmology running the parallel TreeSPH 
code \textsc{Gasoline} \citep{Wadsley04}. Details of the main simulation are given in \citet{Guedes11}, and are 
briefly summarized here. The high-resolution region, 4 comoving Mpc on a side, is embedded in a low-resolution, dark matter-only, periodic box 
of 90 comoving Mpc on a side, and contains 13 million dark matter particles and an equal number of gas particles, for a final dark and gas particle mass of
$m_{\rm DM}=9.8\times 10^4\,\msun$ and $m_{\rm SPH}=2\times 10^4\,\msun$, respectively. The gravitational softening length, $\epsilon_G$,
was fixed to 120 physical pc for all particle species from $z=9$ to the present, and evolved as $1/(1+z)$ from $z=9$ to the
starting redshift of $z=90$. Compton cooling, atomic cooling, and metallicity-dependent radiative cooling at low temperatures \citep{Mashchenko06}
are included. A uniform UV background modifies the ionization and excitation state of the gas and is implemented 
using a modified version of the \citet{Haardt96} spectrum.

\begin{figure}
\centering
\includegraphics[width=0.48\textwidth]{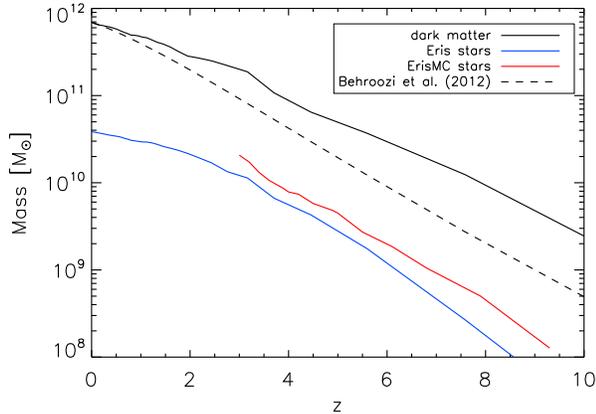}
\vspace{+0.2cm}
\caption{The mass assembly history of ErisMC. The growth of the stellar component ({\it red curve}) is compared to the growth of the 
dark matter halo ({\it black solid curve}). For comparison, we also show the stellar mass in Eris ({\it blue curve}) as well as the best-fit
to the median virial mass for progenitors of halos with mass $M_{\rm vir}=8\times 10^{11}\,\msun$ at $z=0$ in the Bolshoi, MultiDark, and Consuelo
simulations \citep{Behroozi12}. 
}
\label{fig2}
\vspace{+0.1cm}
\end{figure}

Star formation occurs stocastically when cold ($T<3\times 10^4$ K), virialized gas reaches a threshold density $n_{\rm SF}=5$ atoms cm$^{-3}$ and is part
of a converging flow. It proceeds at a rate
\begin{equation}
d\rho_*/dt=0.1 \rho_{\rm gas}/t_{\rm dyn} \propto \rho_{\rm gas}^{1.5}
\label{eq:KS}
\end{equation}
(i.e. locally enforcing a Schmidt law), where $\rho_*$ and $\rho_{\rm gas}$ are the stellar and gas densities, and $t_{\rm dyn}$ is the
local dynamical time. Each star particle has an initial mass $m_*=6\times 10^3\,\msun$ and represents a simple stellar population with its own age, metallicity,
and a \citet{Kroupa93} initial stellar mass function (IMF). Star particles inject energy, mass, and metals back into 
the ISM through Type Ia and Type II SNe and stellar winds,
following the recipes of \citet{Stinson06}. Each SN II deposits metals and a net energy of $0.8 \times 10^{51}\,$ergs into the nearest neighbor gas particles, and 
the heated gas has its cooling shut off (to model the effect of feedback at unresolved scales) until the end of the momentum-conserving phase of the 
SN blastwave, which is set by the local gas density and temperature and by the total amount of energy injected \citep{McKee77}. Cooling is turned off 
only for those particles within the blast radius (for a maximum of 32 neighboring gas particles), and no kinetic energy is explicitly assigned to them. 
For the typical conditions of star-forming clouds resolved in this study, this translates into just one gas particle heated up by a SN and having 
its cooling shut off for a timescale $t_E\sim 5\times 10^5$ yr. The energy injected 
by many SNe adds up to create larger hot bubbles and longer shutoff times. The advantage of this physically-motivated feedback model compared to 
other ``sub-grid" schemes \citep[e.g.][]{Springel03,Oppenheimer08} is that it keeps galactic outflows hydrodynamically coupled to the energy injection by 
SNe (albeit with a delay). In combination with a high gas density threshold for star formation (which enables energy deposition
by SNe within small volumes), this scheme has been found to be key in producing realistic dwarf galaxies \citep{Governato10} and late-type massive spirals   
\citep{Guedes11}.

\begin{figure}
\centering
\includegraphics[width=0.48\textwidth]{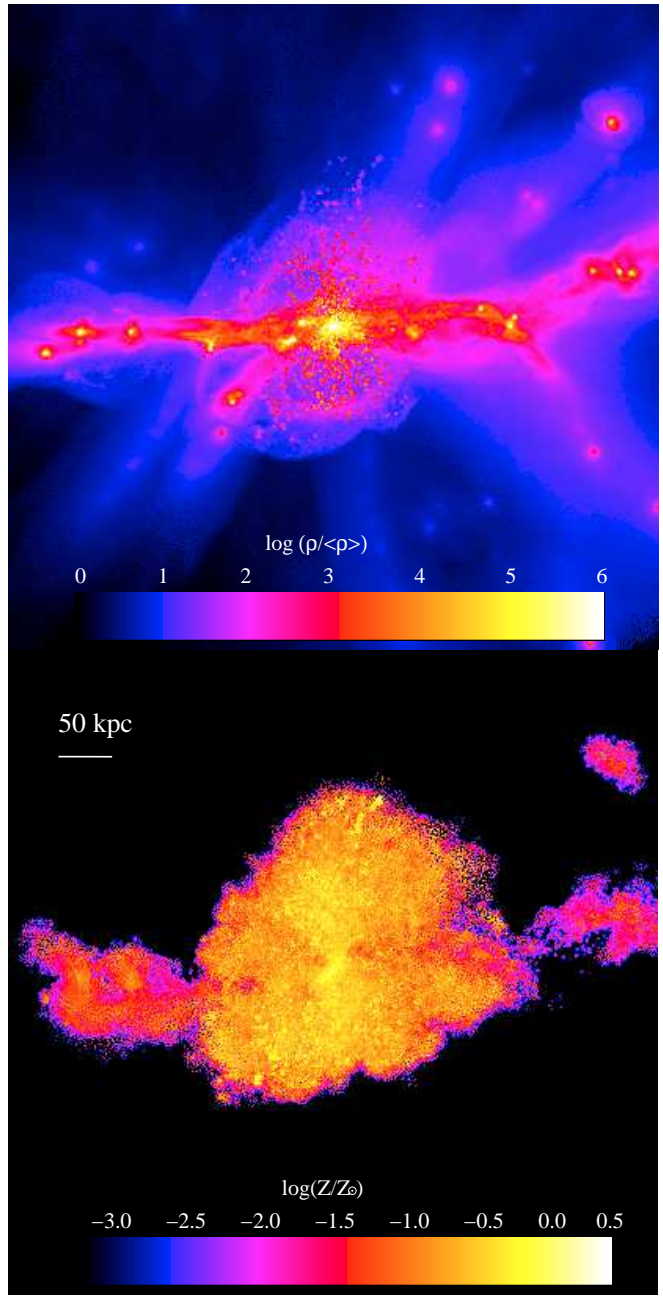}
\vspace{+0.2cm}
\caption{Projected gas density ({\it top panel}) and gas particle metallicity ({\it bottom panel}) of ErisMC's CGM at $z=3$ in 
a cube of 500 (proper) kpc on a side. The galaxy's stellar disk is edge-on in this projection. 
}
\label{fig3}
\vspace{+0.1cm}
\end{figure}

Metal enrichment from SN II and SN Ia follows the model of \citet{Raiteri96}. For SN II, metals are released as the main sequence progenitors die and 
distributed to gas within the blastwave radius. Iron and oxygen are produced according to the following fits to the \citet{Woosley95} yields:   
\begin{equation}                                                                
M_{\rm Fe} = 2.802 \times 10^{-4} \left({m_*\over \msun}\right)^{1.864}\,\msun,                                   
\end{equation}                                                                  
and
\begin{equation}                                                                
M_{\rm O} = 4.586 \times 10^{-4} \left({m_*\over \msun}\right)^{2.721}\,\msun. 
\end{equation}                                                                  
Each SN Ia produces 0.63 $\msun$ of iron and 0.13 $\msun$ of oxygen \citep{Thielemann86} and the metals are distributed between the nearest gas particles. 
Radiative cooling is not disabled following a SN Ia. Stellar wind feedback was implemented based on \citet{Kennicutt94}, and the returned mass fraction 
was determined following \citet{Weidemann87}. The returned gas has the same metallicity as the star particle.   

\begin{figure*}
\centering
\includegraphics[width=0.85\textwidth]{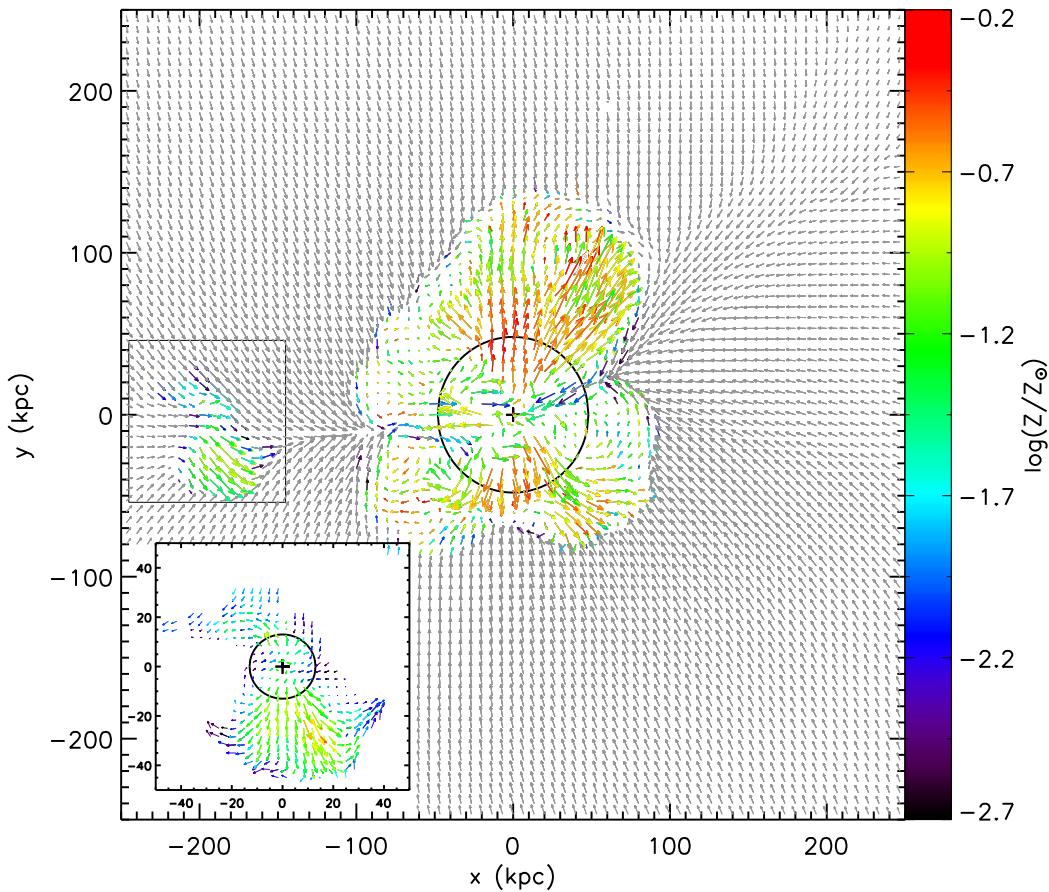}
\vspace{+0.0cm}
\caption{Same as Fig. \ref{fig3}, showing the mass-weighted gas metallicity and velocity field in a $500\times 500\times 10$ kpc$^3$ slice. The galaxy 
center is indicated by the 
plus sign at coordinates $(x,y) = (0,0)$, and its virial radius is marked by the black circle. The color bar indicates the mean gas metallicity within 
the slice, with pristine gas shown in gray. The arrows depict the mass-weighted peculiar velocity vectors of the gas relative to the main host's 
center, with the longest of them representing projected mean velocities of 258 $\kms$ (and 164 $\kms$ in the inset). 
The inset in the bottom left of the panel shows a zoom into the velocity field around one of ErisMC nearby dwarfs ($M_s=5\times
10^9\,\msun$) relative to its own center (indicated by a plus sign in the inset). While being accreted, the dwarf is also venting heavy elements 
into the surroundings. Note that, owing to the averaging process, the metallicities and velocities plotted in this figure can be significantly 
lower than the corresponding quantities for individual gas particles.}
\label{fig4}
\vspace{+0.1cm}
\end{figure*}

The reference simulation discussed here, ``ErisMC", includes metallicity-dependent radiative cooling at high temperatures following \citet{Shen10}.
Aside from high-temperature metal cooling, Eris and ErisMC were run with exactly the same setup. 
   
\section{ErisMC's circumgalactic medium}
\label{metaltrace}

Figure \ref{fig1} shows the optical/UV stellar properties of ErisMC at $z=3$. The mock images were created using the radiation transfer code {\sc 
Sunrise} \citep{Jonsson06}, which produces spectral energy distributions using the age and metallicities of each simulated star particle, and 
takes into account the three-dimensional effect of dust reprocessing. At this epoch, the simulated galaxy has a virial mass of $M_{\rm vir} = 2.4 
\times 10^{11}\,\msun$, a virial radius of $R_{\rm vir} = 48$ kpc, a total stellar mass of $M_*=2.1\times 10^{10}\,\msun$, and is forming stars at the
rate of SFR$=18\,\sfr$.

The mass assembly history of ErisMC's stellar component and dark halo is shown in Figure \ref{fig2}.  While ErisMC is in the right range 
of halo masses ($M_{\rm vir}=10^{11.5\pm 0.3}\,\msun$, \citealt{Adelberger05b}) and stellar masses ($M_*=10^{10.32\pm 0.5}\,\msun$, \citealt{Shapley05}) of 
LBGs at redshifts 2-3, by $z=3$ it has formed 70\% more stars than Eris as a consequence of the increased metallicity-dependent radiative 
cooling at high temperatures. We note here that a recent new simulation of the Eris suite (``Eris2", see \citealt{Shen12}) that includes metal 
diffusion, a \citet{Kroupa01} IMF that boosts the number of Type II supernovae per unit stellar mass by about a factor of 2 compared to \citet{Kroupa93},
and metallicity-dependent radiative cooling at all temperatures, produces by redshift 3 the same stellar mass of Eris. This shows how uncertanties 
in the IMF, metal diffusion and gas cooling properties can affect the star formation history of the simulated galaxy. A comprehensive analysis of 
the average star formation rates and histories of galaxies and their connection to the underlying growth and merging of dark matter
halos has been recently presented by \citet{Behroozi12}, and it is interesting to discuss ErisMC in this context.   
With a specific star formation rate (sSFR=SFR/$M_*$) of $8.6\times 10^{-10}$ yr$^{-1}$ at $z=3$, ErisMC is consistent with the sSFR expected 
at these redshifts for such massive stellar systems (see Fig. 4 of \citealt{Behroozi12}).   
According to \citet{Behroozi12}, halos of mass $\sim 10^{12}\,\msun$ are the most efficient at forming stars at every epoch, 
with baryon conversion efficiencies of 20-40\% that are constant to within a factor of 2 over a remarkably large 
redshift range. ErisMC's efficiency, about 50\%, appears then too high for a ``typical" $M_{\rm vir}=2.4\times 10^{11}\,\msun$ halo at $z=3$.
Figure \ref{fig2} shows, however, that the dark matter accretion history of Eris (the same as ErisMC) is far from 
``typical": when compared to the median virial mass 
for progenitors of halos with mass $8\times 10^{11}\,\msun$ at $z=0$ in the Bolshoi, MultiDark, and Consuelo simulations (see fit in \citealt{Behroozi12}), 
Eris's fractional growth appears to be more skewed towards high redshift. While it is conceivable that this may cause 
Eris (and ErisMC) to form their stars "too early", it is fair to keep in mind that zoom-in hydrodynamical simulations may generically suffer
from having star formation efficiencies that are too high at early epochs, as recently argued by \citet{Moster12}. 

A 2D photometric decomposition performed on the 
dust-reddened rest-frame $i$-band light distribution with the {\sc Galfit} program \citep{Peng02} shows the presence in ErisMC of an extended 
stellar disk with radial scale length $R_d=$ 0.6 kpc. The total gas and stellar metallicities are $Z_g=0.08\,Z_\odot$ and 
$Z_*=0.19\,Z_\odot$, respectively. The galaxy's gaseous disk (defined by all the cold, $T<3\times 10^4$ K gas within 15 comoving kpc from the center) 
is characterized by $12 +\log {\rm (O/H)}=8.4$. This is below the mass metallicity relation at $z=0$ from the {\it Sloan Digital Sky Survey} 
\citep{Tremonti04} but in agreement with the $z\sim 3$ metallicity data on LBGs from \citet{Mannucci09}.

\begin{figure}
\centering
\includegraphics[width=0.48\textwidth]{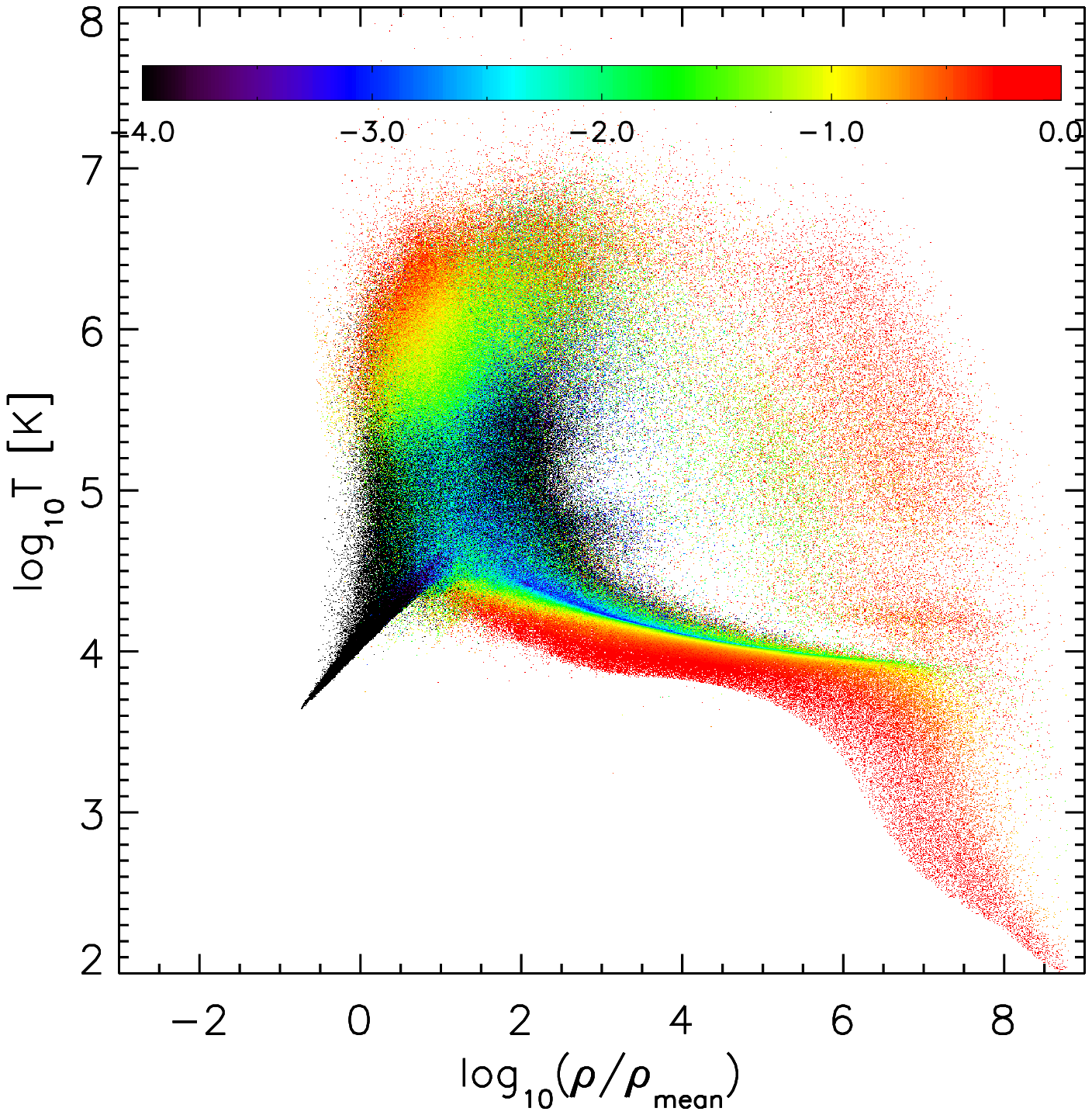}
\vspace{+0.cm}
\caption{Distribution of all enriched gas in the temperature-density plane at $z=3$ within 250 physical kpc from the center of the main host. 
The color scale indicates the mass-weighted metallicity. 
}
\label{fig5}
\vspace{+0.1cm}
\end{figure}

The complex environment of the galaxy forming region is clearly seen in Figure \ref{fig3}, which shows projected gas density (top panel) and 
gas-phase metallicity (bottom panel) in a cube 500 kpc on a side. While metal-enriched material is seen as far as 200 (proper) kpc (i.e. more than $4R_{\rm vir}$) from 
the center, within the same region we also identify 47 self-bound  dwarfs galaxies with masses $M_s>10^8\,\msun$ (424 with $M_s>10^7\,\msun$), many of them also 
forming stars and polluting their surroundings. The total mass of heavy elements in the gas phase within $R_{\rm vir}$, $2R_{\rm vir}$, and $3R_{\rm vir}$ 
is $1.7\times 10^7\,\msun$, $2.9\times 10^7\,\msun$, and $3.0\times 10^7\,\msun$, respectively. A region $\sim 100$ kpc in size is enriched to metallicities 
above 0.03 solar. The extent of the metal enriched region is consistent with recent observations of circumgalactic metals around LBGs \citep{Steidel10}.

\begin{figure}
\centering
\includegraphics[width=0.48\textwidth]{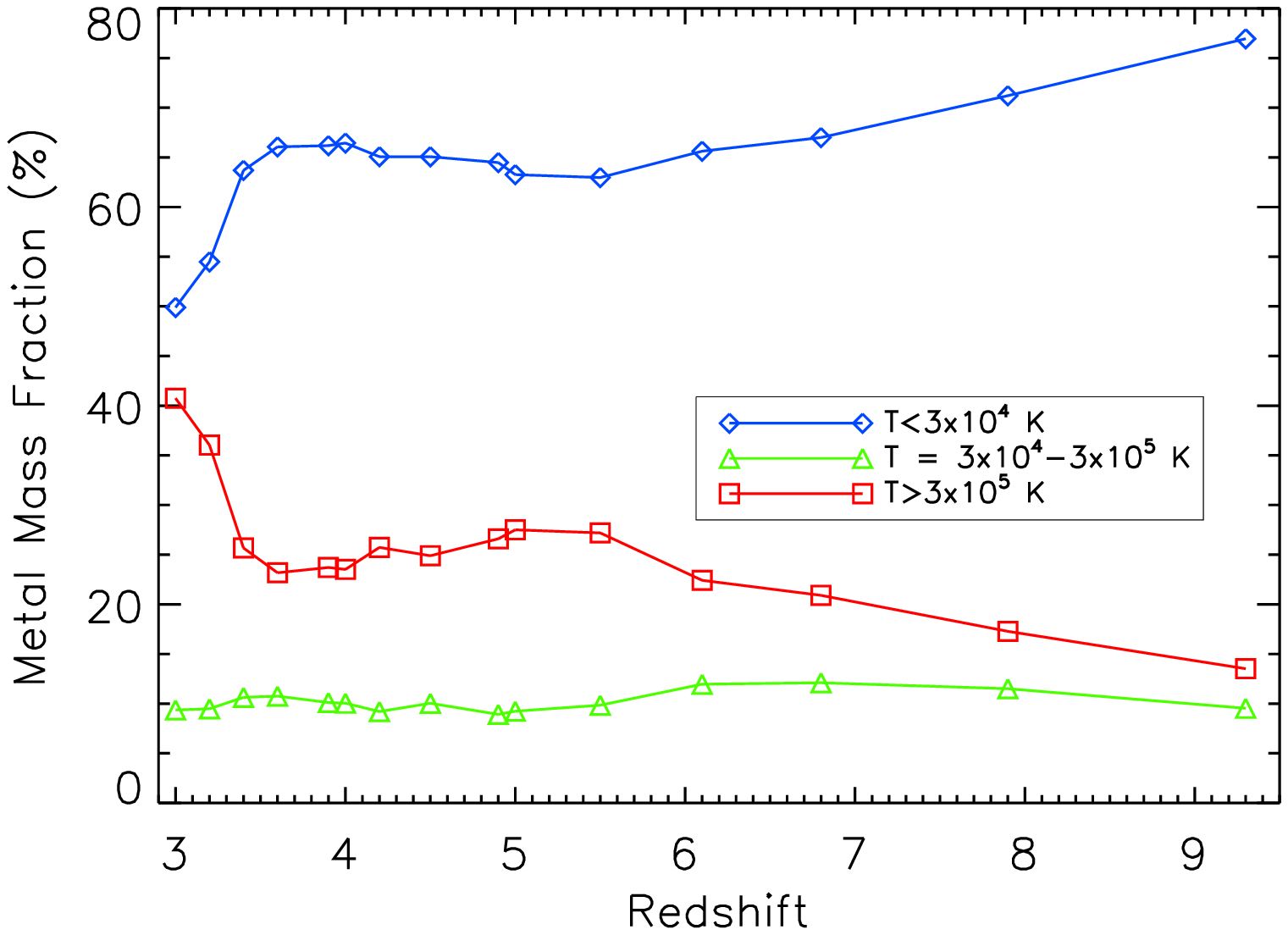}
\includegraphics[width=0.48\textwidth]{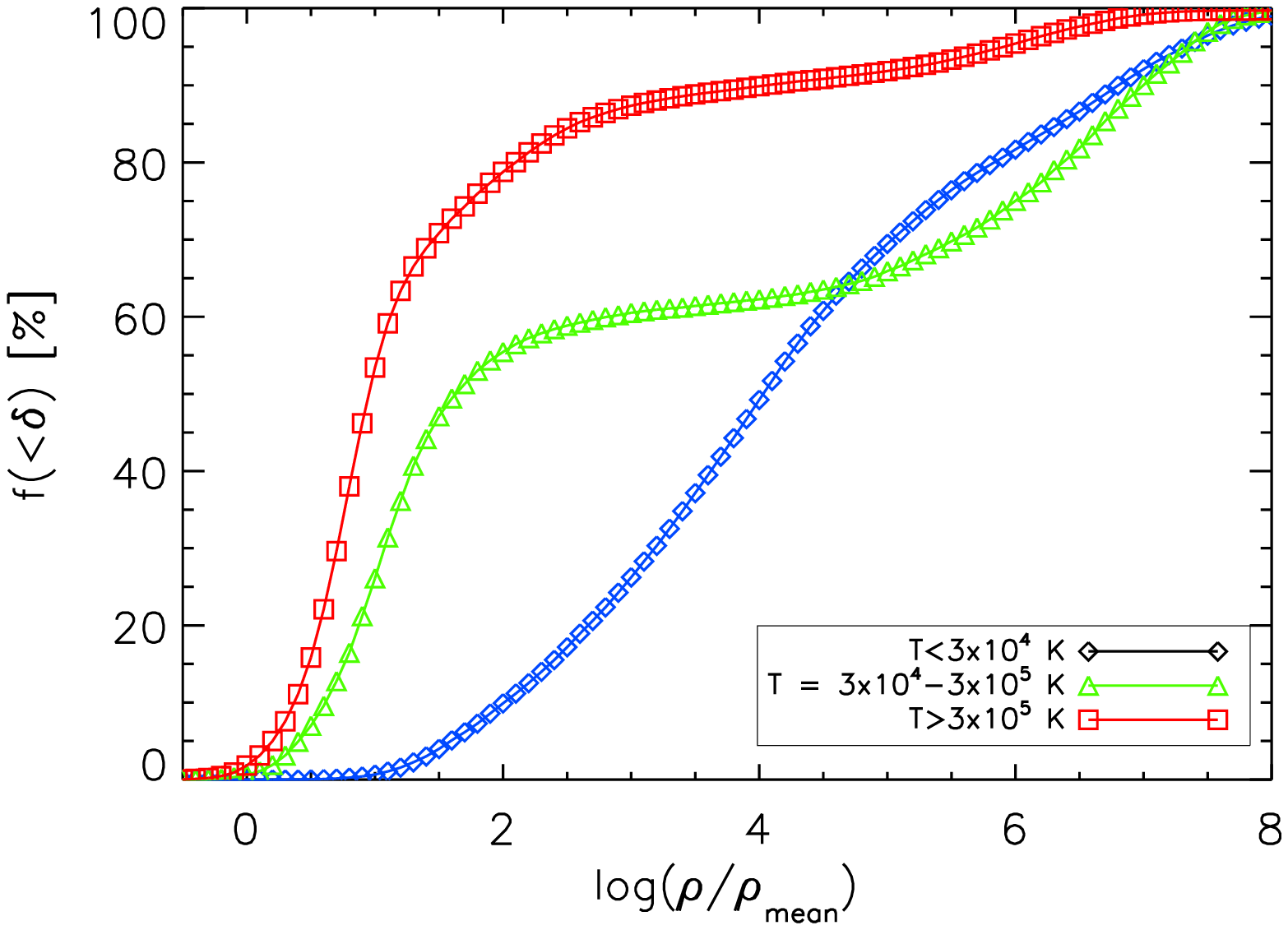}
\vspace{+0.cm}
\caption{{\it Top panel:} Evolution with redshift of the mass fraction of heavy elements in the gas phase within 1 comoving Mpc (250 physical kpc at z = 3) from the center of the main host.
The three curves shows the fraction of metals in the cold ($T<3\times 10^4$ K), warm ($3\times 10^4 {\rm K}<T<3\times 10^5$ K), and hot ($T>3\times 10^5$) 
interstellar and circumgalactic gas.  {\it Bottom panel:} Cumulative metal mass fraction in each gas phase at $z=3$ as a 
function of overdensity.
}
\label{fig6}
\vspace{+0.cm}
\end{figure}

Figure \ref{fig4} shows a projected gas metallicity and velocity map
in a $500\times 500\times 10$ kpc$^3$ slice. The arrows indicate the direction and magnitude of the mass-weighted peculiar velocity field relative to the center of ErisMC.
Galactic winds can be clearly seen propagating perpendicularly to the disk (seen edge-on in this projection) well beyond the virial radius, with average velocities (over the
slice) that exceed $250\,\kms$ (we shall see below that individual gas particle speeds can reach 800 $\kms$). The outflows have a bipolar 
distribution with a smaller opening angle near the base, similar to the observed morphologies of galactic winds in star-forming 
galaxies \citep{Veilleux05}. Inflows along large-scale filaments bring in both pristine gas (gray arrows in the figure), as well as material pre-enriched by 
nearby dwarfs (colored arrows on the left side of the figure and in the inset). Inflowing cold streams that penetrate deep inside the virial radius  
are commonly seen in cosmological hydrodynamical simulations \citep[e.g.][]{Keres05,Ocvirk08,Dekel09}, and have been shown to give origin to 
\CII\ absorption with a significant covering factor \citep{Shen12}. 

\begin{figure*}
\centering
\includegraphics[width=0.85\textwidth]{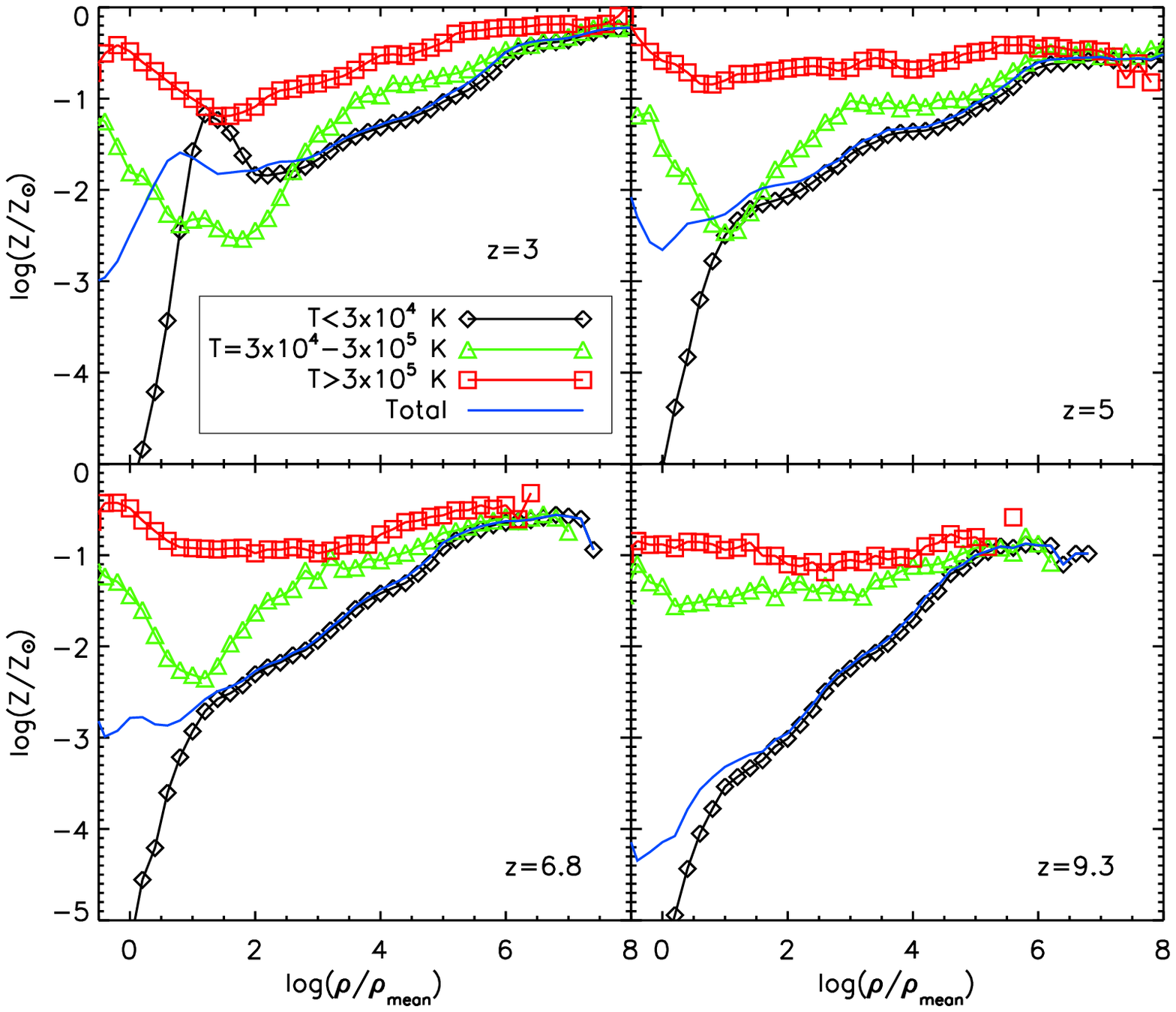}
\vspace{+0.2cm}
\caption{Mean metallicity of cold, warm, and hot gas within 1 comoving Mpc (250 physical kpc at z = 3) from ErisMC's center as a function of gas overdensity in four redshift bins. 
The thin blue line shows the total gas metallicity.
}
\label{fig7}
\vspace{+0.2cm}
\end{figure*}

It is instructive at this stage to look at the mass-weighted distribution of all (inside and outside the main host) enriched gas at $z=3$
in the temperature-density plane.
Figure \ref{fig5} indicates that metals are spread over a large range of phases, from cold star-forming material at $T<3\times 10^4$ K and $n>n_{\rm SF}=5$
atoms cm$^{-3}$ (corresponding to $\delta\equiv \rho/\rho_{\rm mean}>3\times 10^5$ at $z=3$) to hot $T>10^6$ K low density $\delta=3$ intergalactic gas
that cannot cool radiatively over a Hubble time.
The black strip in the lower left corner of the figure marks the pristine, adiabatically cooling IGM, while the colored swath in the
lower right corner shows dense, metal-rich gas in the galaxy disk cooling down below $10^4$ K.
Hot enriched gas vented out in the halo by the cumulative effect of SN explosions can be seen cooling and raining back 
onto the disk in a galactic fountain. Intergalactic gas in the range $1\lta \delta \lta 10$ shows a strong positive gradient of 
metallicity with increasing temperature, and has the largest range of metallicities, extending from solar all the way down to zero. 
We note that this is the temperature-density plane for gas in a limited zoom-in region surrounding the main galaxy, and that these results 
may not be representative of the $z=3$ IGM as a whole and of the CGM of other galaxies. 

\begin{figure}
\centering
\includegraphics[width=0.49\textwidth]{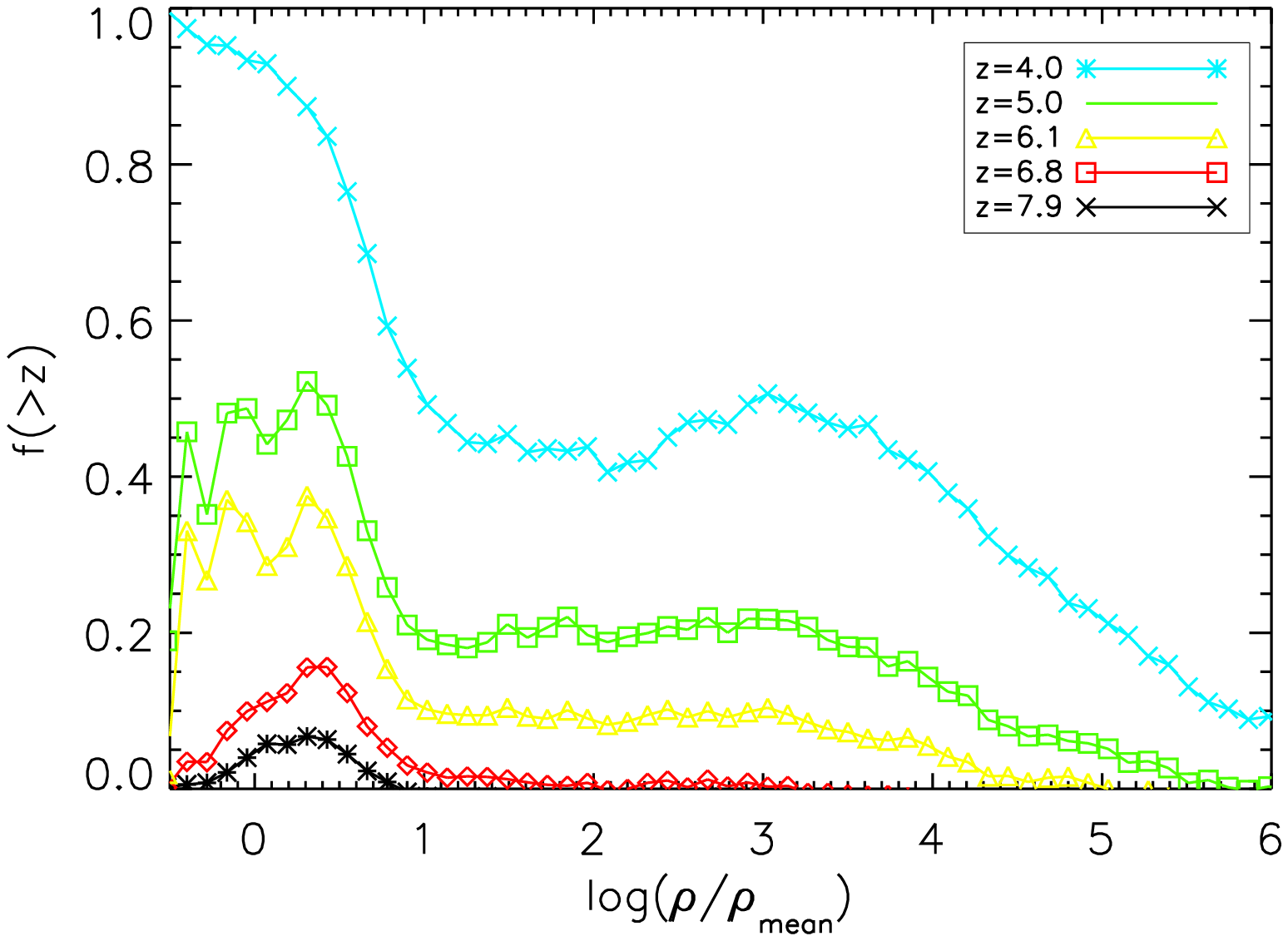}
\vspace{+0.0cm}
\caption{Fractional amount of metals within 250 physical kpc from ErisMC's center at redshift 3 at overdensity $\delta(z=3)$ that was added to gas particles before redshift $z=4,5,6.1,6.8,7.9$.
}
\label{fig8}
\vspace{0.1cm}
\end{figure}

\begin{figure*}
\centering
\includegraphics[width=0.85\textwidth]{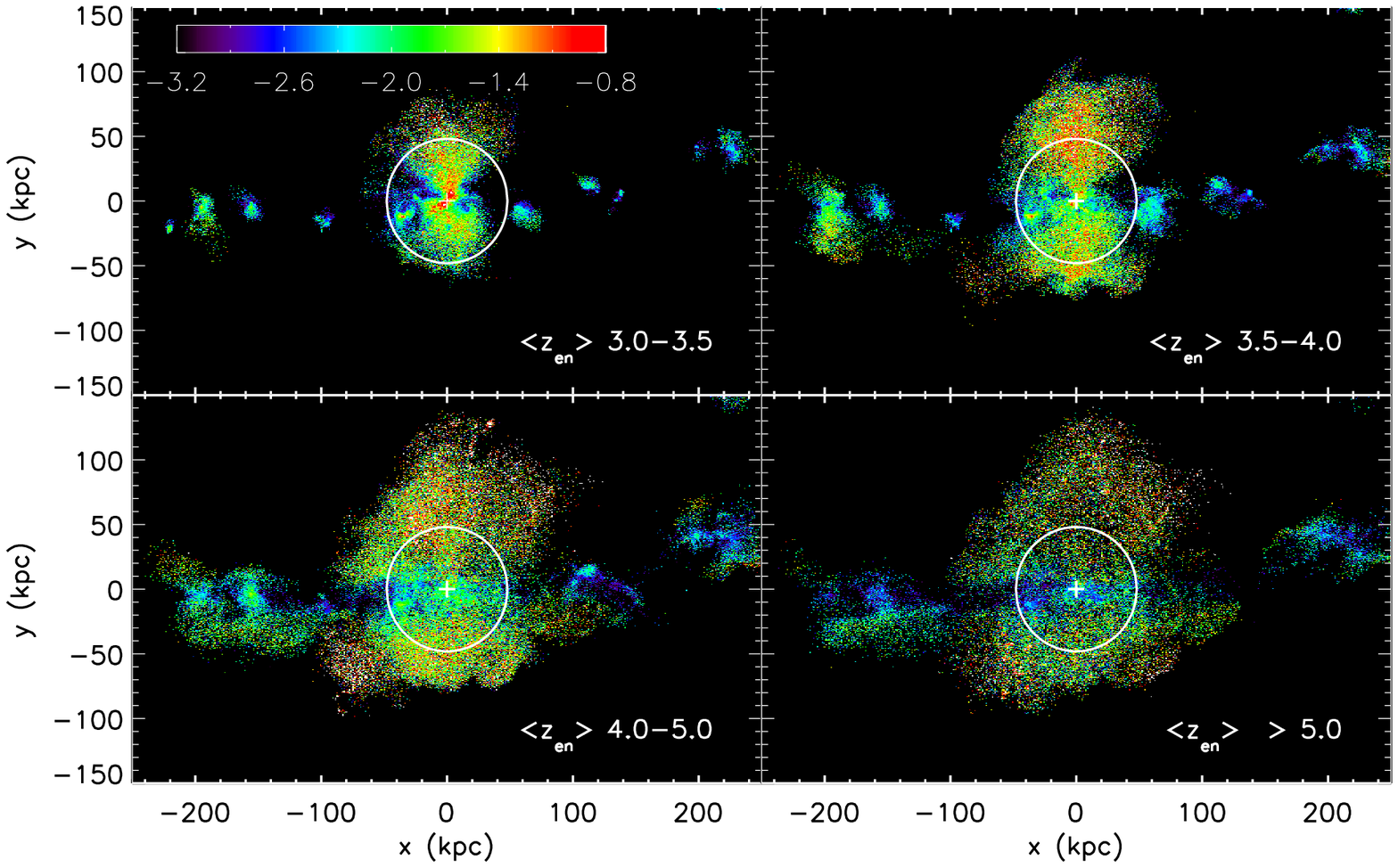}
\vspace{+0.0cm}
\caption{Projected gas metallicity map of ErisMC's CGM at $z=3$ in a $500\times 300 \times 300$ kpc$^3$ (proper) box. The galaxy center and its
virial radius are indicated by the plus sign and the black circle, respectively. The color coding indicates the mean gas metallicity in a column 
of area $dxdy=7.1\times 4.3$ kpc$^2$. The different panels highlight different enrichment redshift ranges. 
Gas enriched at earlier epochs generally lies at larger distance to the main host's center, as galactic outflows transport metals from the dense regions 
of star formation into the CGM on cosmological timescales. At galactocentric distances $> 130$ kpc the IGM is enriched mostly by nearby dwarfs.
} 
\label{fig9}
\vspace{+0.1cm}
\end{figure*}

A census of all the gas-phase metals in the cold ($T<3\times 10^4$ K), warm ($3\times 10^4 {\rm K}<T<3\times 10^5$ K), and hot ($T>3\times 10^5$ K) 
interstellar and circumgalactic medium within 1 comoving Mpc (250 physical kpc at redshift 3) from the center of ErisMC is depicted in Figure \ref{fig6} (top panel) as a function of redshift. 
The fraction of metals in the cold gas drops from about 70\% at redshift 8 to 50\% at $z=3$. Hot gas is the second most important reservoir of 
gas-phase heavy elements, with a fraction of metals that increases from 15\% at $z=8$ to 40\% at $z=3$. This phase would remain 
undetected in UV spectroscopic studies of high redshift galaxies, and would therefore contribute to the ``missing metals" 
\citep{Pettini06,Bouche07}. The metal mass fraction in warm gas remains around 10\% at all epochs. Note that the sudden increase  
in the amount of hot metals at $z<4$ is caused by vigorous outflow activity at these epochs rather than by the transition from ``cold" to ``hot" 
accretion mode in ErisMC. 

More than 40\% of all gas-phase metals at $z=3$ lie outside the virial radius: while cold metal-rich material traces large overdensities 
within the main host, about 50\% of all warm and 70\% of all hot metals are found in low density $\delta<30$ regions beyond the virial radius, 
a point illustrated in the bottom panel of Figure \ref{fig6}. Intergalactic metals are characterized by a strong temperature gradient with overdensity,
as the metal-weighted temperature climbs from $10^4$ K at $\delta=1$ to $2\times 10^5$ K at $\delta=10$.     

The metallicities of the individual cold, warm, and hot gas components, as well as that of the total gas phase, are shown in Figure \ref{fig7} as a
function of gas overdensity in four redshift bins. The gas-phase metallicity distribution at $z=3$ shows a positive density gradient 
above an overdensity of $\log\delta=2$, with little redshift evolution as heavy elements removed from high density regions are steadily replenished. 
The mean total metallicity exhibits a plateau around $\log \delta=1-2$, where the metals transported by SN-driven winds accumulate, and 
drops quickly below $\log Z/Z_\odot=-2$ at overdensities below a few. A low-density peak is more clearly seen in the metallicity distribution 
of cold, warm, and hot gas: in the case the warm and hot medium the peak is shifted towards underdense $\delta \lta 1$ regions.
At $z=3$, the mean metallicity of cold gas drops quickly below $\log Z/Z_\odot=-2$ at overdensities $\delta\lta 10$. 

\begin{figure}
\centering
\includegraphics[width=0.49\textwidth]{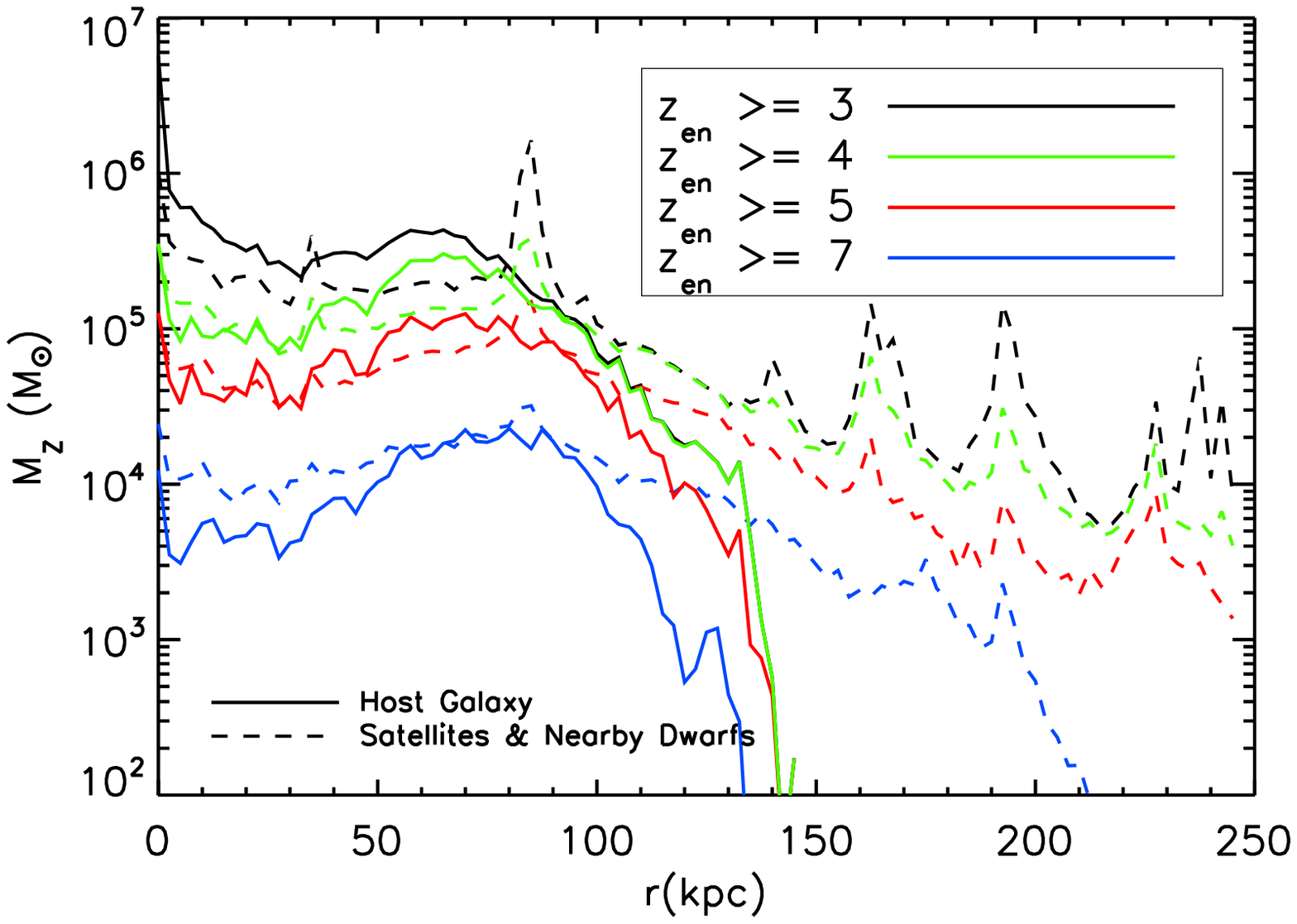}
\caption{Gaseous metal mass in 2.5 kpc thick radial shells at varying distances from the center at z = 3.  These metals were released by $z=3$ from ErisMC's 
main halo ({\it solid lines}) and its satellites and nearby dwarfs ({\it dashed lines}). The colors indicate metals produced at different enrichment epochs.
}
\label{fig10}
\end{figure}

\begin{figure*}
\centering
\includegraphics[width=0.85\textwidth]{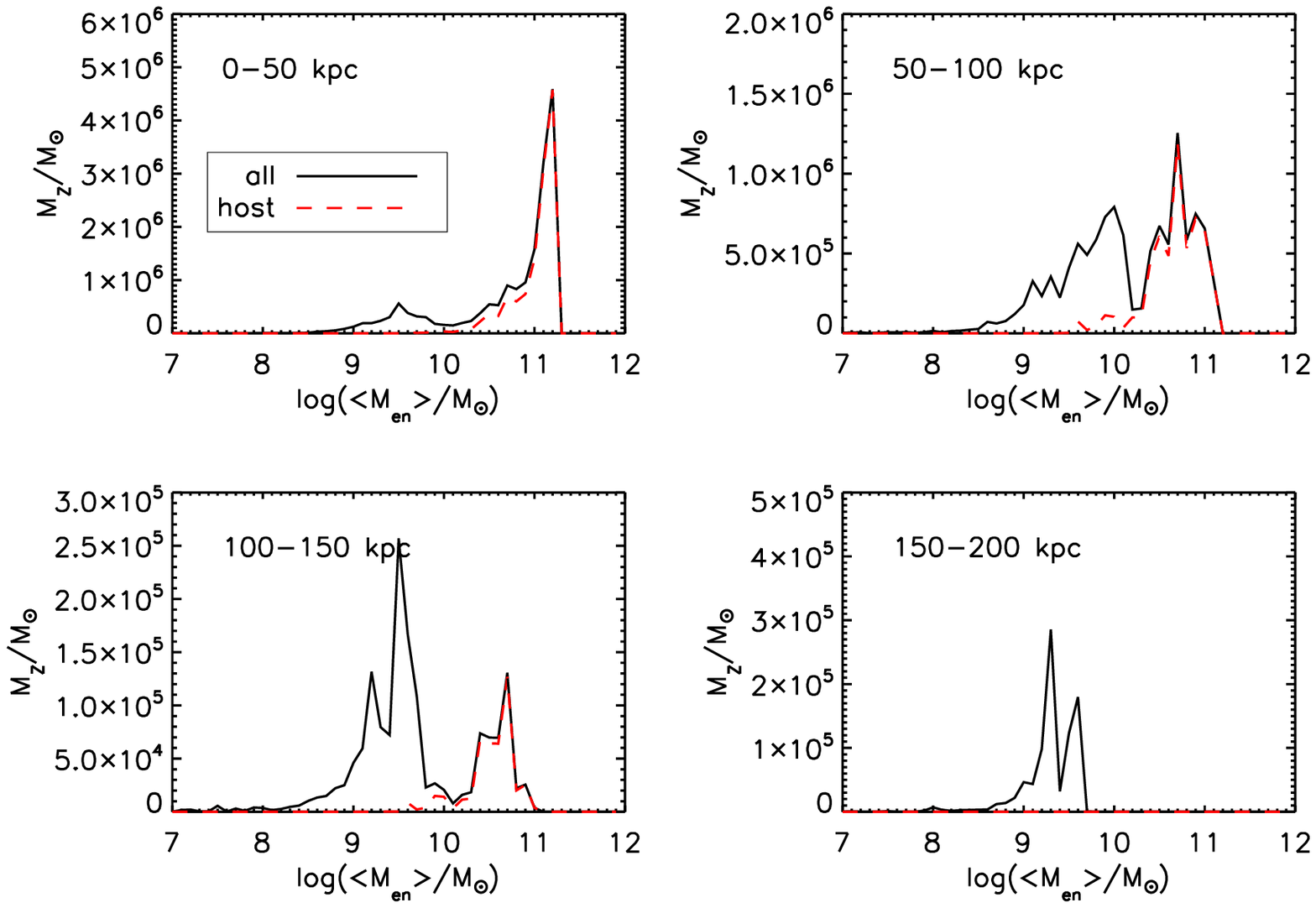}
\caption{Metal mass in ErisMC's CGM as a function of the mean source halo mass $\langle M_{\rm en}\rangle$. Each panel indicates 
the physical distance of the gas from the main halo's center at $z=3$. {\it Solid lines:} main halo plus satellites and nearby dwarfs. {\it Dashed lines:} main 
halo only.  The mass bin in all panels is $\Delta (\log \langle M_{\rm en}\rangle)=0.1$.
}
\label{fig11}
\end{figure*}

While our ``zoom-in" simulation focuses on the CGM of a single massive system and cannot describe the enrichment history of 
the IGM as a whole by a population of galaxies with different masses and star formation histories, it is still instructive to briefly 
compare our results with those of very recent, much lower resolution, large cosmological volume simulations of IGM pollution that use 
different sub-grid prescriptions for generating galactic outflows \citep{Wiersma09,Cen11,Oppenheimer11}. \citet{Cen11} inject thermal energy 
and metals from SN feedback on tens of kiloparsecs scales, do not turn off hydrodynamic coupling between the ejected metals and the surrounding gas, and, 
like in ErisMC, find that a large fraction of all the heavy elements in the gas phase at redshift 3 have temperatures in excess of $3\times 10^4$ K. 
Their metallicity-density relation for cold gas also shows a low-density peak, albeit shifted towards underdense regions compared to ErisMC.  
For comparison, in the momentum-conserved wind implementation of kinetic feedback by \citet{Oppenheimer11} (where hydrodynamic forces 
are temporarily turned off) there is no low-density peak in the metallicity-density relation and metals largely reside in cool gas. IGM metals 
reside primarily in a warm-hot component in the simulations of \citet{Wiersma09}, who also use kinetic feedback but with non-decoupled wind models.
Measurements of the distribution of carbon in the IGM using pixel statistics yield, at $z=3$ and $\log\delta=0.5$, [C/H]$\approx -2.8$ \citep{Schaye03}, 
a value that depends on the hardness of the UV background. In ErisMC we measure, for gas within 250 physical kpc from the center of the host, and at the same overdensity and including only cold and warm gas, [C/H] is also around $-2.8$. 

\section{The origin of circumgalactic metals}

In the absence of metal diffusion \citep{Shen10}, the SPH technique allows us to trace back in time the enrichment history of every gas particle. 
In this and the following section we use simulation outputs from the ErisMC run and the Amiga's Halo Finder (AHF, \citealt{Knollmann09}) 
to identify the age of the metals observed at redshift 3 and the sources of pollution -- whether nearby dwarfs, satellite progenitors, or 
the main host.  

Figure \ref{fig8} shows the fraction of metals at redshift 3 that was released to gas particles at overdensity $\delta(z=3)$ before 
redshifts $z=4,5,6.1,6.8,7.9$. The epoch at which a gas element is enriched is clearly a sensitive function of its overdensity at some later time. 
The diffuse IGM is typically enriched earlier than high density regions, a trend that is in agreement with the results of \citet{Wiersma10}
and \citet{Oppenheimer11}. More than 50\% (35\%) of all $z=3$ metals at the average density were synthesized before $z=5$ ($z=6$), 
while newly produced metals are mostly confined to high overdensities.

\begin{figure*}
\centering
\includegraphics[width=0.85\textwidth]{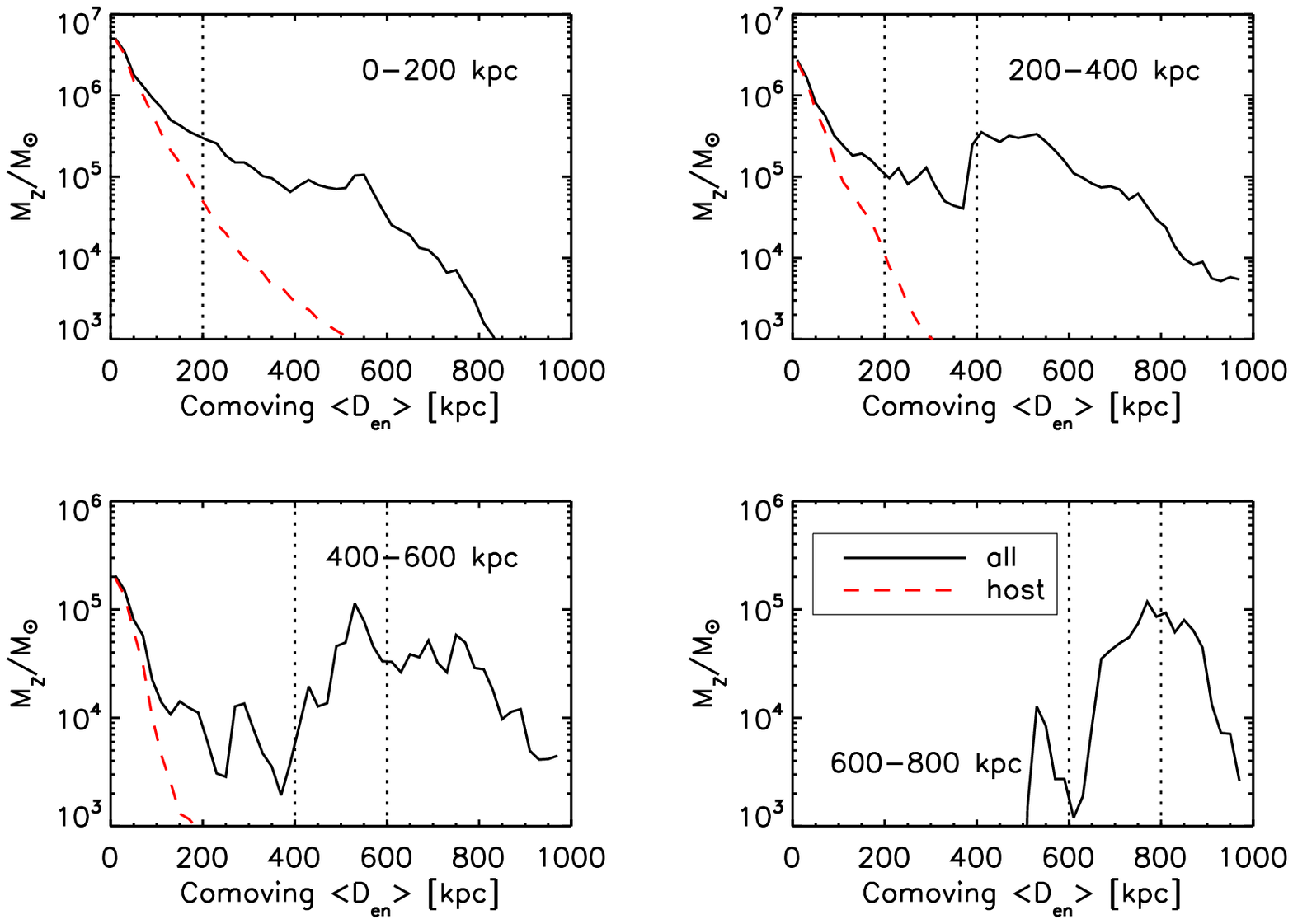}
\caption{Metal mass distribution at $z=3$ as a function of the comoving mean enrichment distance $\langle D_{\rm en}\rangle$. The dotted vertical lines and the legend 
in each panel mark the comoving distance interval from ErisMC's center of the enriched gas under consideration at z = 3.  {\it Solid lines:} main halo 
plus satellites and nearby dwarfs. {\it Dashed lines:} main host only.}
\label{fig12}
\end{figure*}

There are two possible causes for this ``outside-in" (a term we borrow from \citealt{Oppenheimer11}) enrichment of ErisMC's CGM: 1) 
metals released at earlier times into the high-density regions of ErisMC main host and transported into the IGM via galactic winds on timescales that are 
comparable to the age of the universe at $z =3$; and 2) metals released at earlier times in dwarf galaxies and shed into the surrounding intergalactic 
and circumgalactic medium before and/or during infall. Dwarf galaxies are referred to as ``nearby dwarfs" 
if they are still orbiting outside $R_{\rm vir}$ at $z=3$, and as ``satellite progenitors" if they have been accreted by the main 
host before redshift 3. To identify the time, location, and source of enrichment of a given gas particle, we define a metal mass-weighted redshift 
as in \citet{Wiersma10}: 
\begin{equation}
\langle z_{\rm en}\rangle = \frac{\sum_{i} \Delta m_{Z,i} z_i}{\sum_{i} \Delta m_{Z,i}},
\label{eqn:zen}
\end{equation}
where $\Delta m_{Z,i}$ is the metal mass gained by the gas particle in an enrichment event at redshift $z_i \ge 3$, and $\sum_i \Delta m_{Z,i}$ 
is the total metal mass of the particle at $z=3$. Gas particles in ErisMC typically receive metals more than once, with about half of them enriched in more than 
three events. For every enrichment episode we also derive the mass of the satellite or nearby dwarf in which the gas 
particle resides, $M_{h,i}$ and the distance between the gas particle 
and the center of ErisMC, $d_{i}$. Analogously to equation (\ref{eqn:zen}), we then define a metal mass-weighted source halo mass and 
distance from ErisMC center as 
\begin{equation}
\langle M_{\rm en}\rangle = \frac{\sum_{i} \Delta m_{Z,i} M_{h,i}}{\sum_{i} \Delta m_{Z,i}} 
\label{eqn:Mhalo}
\end{equation} 
and 
\begin{equation}
\langle D_{\rm en} \rangle = \frac{\sum_{i} \Delta m_{Z,i} d_{i}}{\sum_{i} \Delta m_{Z,i}}. 
\label{eqn:dis}
\end{equation}
respectively. We have separated gas particles into different groups according to their $\langle z_{\rm en}\rangle$ and plotted in Figure \ref{fig9} 
the projected metallicity of each group at redshift 3. Metals within ErisMC's virial radius are clearly ``younger", i.e. they are characterized by an 
enrichment redshift $\langle z_{\rm en}\rangle$ between 3 and 3.5. Because of the long wind 
propagation time, ``older" metals with $\langle z_{\rm en}\rangle$ between 4 and 5 are spread over 100 
(physical) kpc perpendicularly to ErisMC's disk. Low-metallicity gas in this enrichment redshift range can be seen as far as 200 kpc from the main host
center as it is ejected from nearby dwarfs (see also Figs. \ref{fig3} and \ref{fig4}). There is little material contaminated by 
metals at $\langle z_{\rm en}\rangle >5$ within ErisMC's virial radius. Late ($\langle z_{\rm en}\rangle <5$) galactic ``superwinds" -- the result of recent 
star formation in ErisMC's main host -- are found to account for less than 9\% of all the metals observed beyond $2\rvir$,   

\begin{figure*}
\centering
\includegraphics[width=0.85\textwidth]{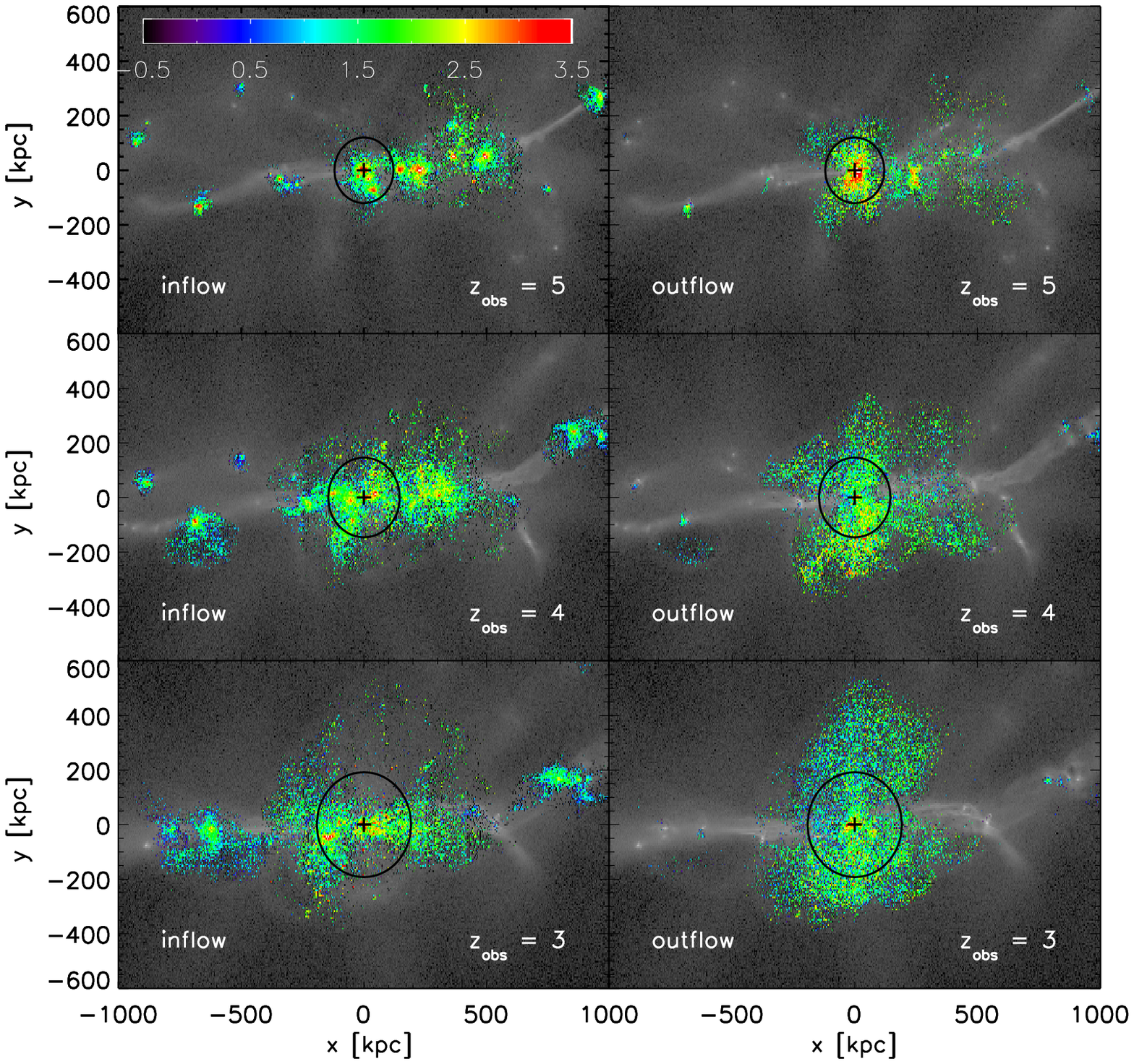}
\caption{Metal column density map of ErisMC's circumgalactic environment at three different observer redshifts, $z_{\rm obs}=3,4,5$. 
All the heavy elements shown in the figure are produced by satellites and nearby dwarfs at epochs $5\le z\le 7$. The projected region has a comoving size 
of 2 $\times$ 1.2 $\times$ 1.2 Mpc$^3$ and is centered on the main host. The center of the host galaxy and its virial radius are indicated by the plus sign and 
black circle, respectively. The color scale indicates the logarithmic of total metal mass a column $dxdy=28.6\times 17.1$ kpc$^2$ comoving, in units of $\msun$. 
{\it Left panels}: inflowing material. {\it Right panels}: outflowing material. 
}  
\label{fig13}
\end{figure*}

Figure \ref{fig10} sheds light on the role played by satellites and nearby dwarfs in contaminating ErisMC's circumgalactic 
medium. It shows the total mass of heavy elements released by the main host and its satellites as a function of distance from ErisMC's center 
at redshift 3. About 60\% of all the metals within 100 kpc of the center originate from the main host, and the rest from its satellites. 
Beyond 100 kpc, nearby dwarfs start dominating the metal budget. Both the host and the satellites contribute to the recent ($z<4$) pollution of gas within the virial radius. 
Older metals within $R_{\rm vir}$ typically form in satellite progenitors, collect along the filaments into the main host, and are not blown away by galactic outflows.
Note how, within 150 kpc or so, the distribution of metals from satellites is rather smooth and follows that from the main host, 
an indication that gas polluted by star formation in satellite progenitors is stirred up and well mixed with ErisMC's galactic outflows.
Spikes due to individual nearby dwarfs can be seen beyond 85 kpc.

It is interesting at this stage to look at the masses of the satellites that contribute to the enrichment of the CGM. Figure \ref{fig11} shows
the metal-weighted mean halo mass, $\langle M_{\rm en}\rangle$ (defined in eq. \ref{eqn:Mhalo}), for gas at different physical distances from ErisMC's center. 
Most of the satellites' metals come from systems more massive than $10^9\,\msun$, with the peak of the distribution typically around $10^{9.5}\,\msun$.   
Satellites smaller than $10^{8.5}\,\msun$ do not cause significant pollution as they are unable to form many stars both because of SN feedback and the
suppression of baryonic infall by the UV background.
Also, halos smaller than $10^{8}\,\msun$ are resolved by less than 1,000 particles in our simulation, and the inability to properly resolve high density 
star-forming regions in these objects may result in a numerical suppression of star formation (as our recipe is based on a high gas density threshold).
The right bottom panel shows how gas beyond 150 kpc is enriched only by nearby dwarfs. Figure \ref{fig12} provides some insight on the overall 
transport of enriched gas by showing the metal mass distribution as a function of the enrichment distance (as defined in eq. \ref{eqn:dis}).
We use comoving distances to highlight departures from pure Hubble flow. Most host metals  are released within 50-100 comoving 
kpc from the center, with those found beyond the virial radius at $z=3$ originating earlier from strong galactic winds launched 
closer to the center. Metals from satellites can have very different kinematics. Some are ejected directly into the IGM by nearby dwarfs, 
other, initially deposited in the halo of the main host by infalling dwarf progenitors, become part of ErisMC's galactic outflows.      
We find that about 40-50\% of all the satellite-produced metals found in a given distance intervals at $z=3$  were produced at larger distances and 
then transported inwards to their current location. For example, 42\% of all satellite metals found at $z=3$ in the range 400-600 (comoving) kpc from ErisMC's center 
have a mean enrichment distance larger than 600 kpc. All metals found beyond 600 comoving (150 physical) kpc were ejected by nearby dwarfs
that have not been yet accreted by the host. 

To better grasp the kinematics of the material enriched by satellite progenitors (i.e. dwarf galaxies that have all been 
accreted by the main halo before $z=3$), we plot in Figure \ref{fig13} a metal column density map of ErisMC's circumgalactic environment 
at three different observer redshifts, $z_{\rm obs}=3,4,5$. {\it All the heavy elements shown in the figure were produced by satellite progenitors 
and nearby dwarfs at epochs $5\le z\le 7$.} All distances are comoving and the metals are separated according to their radial peculiar velocities relative
to the center of the main host, $v_r$, in inflowing ($v_r<0$, left panels) and outflowing ($v_r>0$, right panels). The projected mass 
distribution is shown in the gray scale. Inflowing metals at $z_{\rm obs}=5$ have contaminated the vicinity of their satellite hosts, and are now 
being accreted onto the main host along the filamentary structure. They are subsequently dispersed during the infall and tidal disruption of their 
satellite hosts, become entrained in the galactic wind of the main host, and are ultimately ejected in a bipolar outflow.   

\section{Properties of galactic outflows}
\label{kinematics}

In this section, we study the properties of ErisMC's galactic ouflows and compare them with the observations as well as with other galactic wind models adopted
in cosmological simulations. 

\subsection{Outflow velocity}

\begin{figure}
\centering
\includegraphics[width=0.48\textwidth]{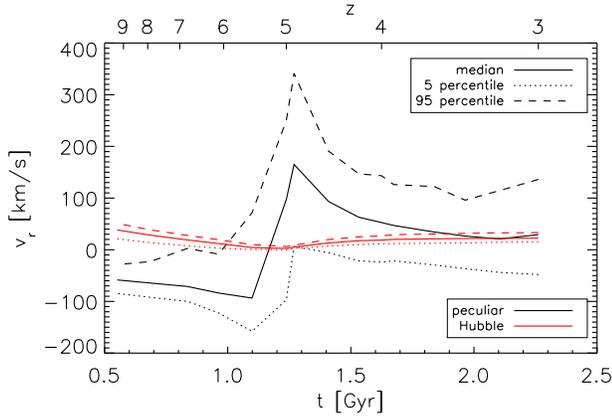}
\caption{Evolution of the radial velocity of gas particles last enriched at redshift 5 in the main host. Peculiar and Hubble flow velocities are 
indicated with the black and red curves, respectively. The solid, dotted, and dashed lines show the median, 5 percentile and 95 percentile values, 
respectively.}
\label{fig14}
\end{figure}

An example of the evolution with time of the radial velocity of polluted material is shown in Figure \ref{fig14}, where we have selected gas particles 
that were enriched for the last time in the main host at redshift 5, and that are {\it unbound} at $z=3$. Gas accretes with negative peculiar velocity
onto the host, is enriched and accelerated to outflow velocities of a few hundred $\kms$ by blastwave feedback, and slows down as it sweeps the ISM and halo 
material. The peak velocity is comparable to those observed in high redshift LBGs \citep[e,g.,][]{Adelberger03,Veilleux05,Steidel10}. It is also 
comparable to the velocities adopted in various kinetic feedback models \citep[e.g.,][]{Springel03,Oppenheimer06,DallaVecchia08,Wiersma09,Choi11}. After reaching a 
peak value, the gas' proper velocity declines rapidly, and particles are swept by the Hubble flow as they move farther from the center into the IGM. This behaviour 
is typical of all enriched gas particles. The rapid decline of the proper velocity is also seen in the kinetic feedback model of \citet{DallaVecchia08}, 
where outflowing particles are allowed to interact hydrodyamically with the ISM, and pressure forces within the disk significantly decrease the wind speed.

\begin{figure}
\centering
\includegraphics[width=0.48\textwidth]{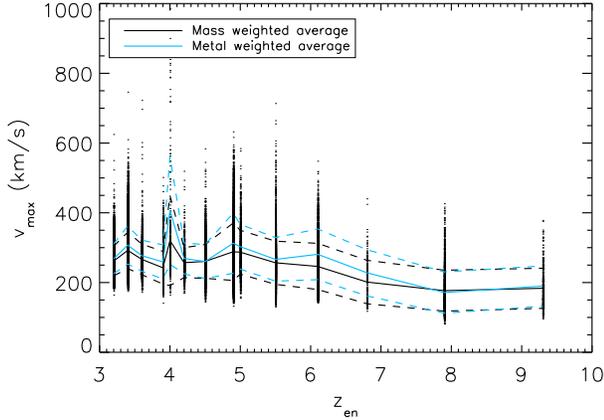}
\caption{Peak radial velocity of gas particles that become unbound after they are enriched in the main host, as a function of their 
enrichment redshift. Each dot represents a gas particle. The black and blue solid lines show the mass-weighted and metal-weighted average values, 
respectively. The colored dashed lines show the standard deviation from the mean. 
}
\label{fig15}
\vspace{+0.2cm}
\end{figure}

\begin{figure}
\centering
\includegraphics[width = 0.48\textwidth]{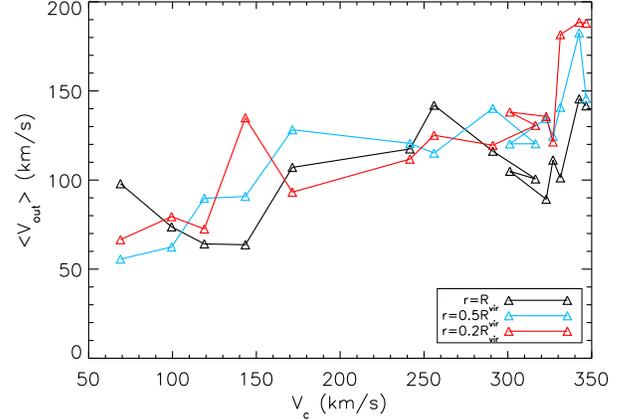}
\caption{Evolution of the mean outflow velocity of all unbound gas particles at different radii for the main host, as a function of its peak circular velocity.
}
\label{fig16}
\end{figure}

Observations of local starburst galaxies \citep[e.g.,][]{Schwartz04, Rupke05} have shown that outflow speeds are correlated with halo masses and star 
formation rates, and that the correlation flattens out for SFR$>10\,\sfr$ \citep{Rupke05}. To investigate the existence of such relationship in the ErisMC
simulation, we plot in the upper panel of Figure \ref{fig15} the peak velocity of gas particles that become unbound after they are enriched in the main host, as 
a function of their enrichment redshift. We choose the gas peak velocity as this is close to the flow speed when the wind is launched, and is 
relatively unaffected by interactions between the outflow and the ISM/gaseous halo. Note that, because of the limited time resolution, 
we may actually underestimate the peak velocity of gas particles that reach a maximum speed and then slow down significantly within the time 
interval of two simulation outputs ($\Delta t \gta 140$ Myr). The scatter plot shows that enriched gas particles can often reach velocities in excess of 
$600\,\kms$, with a few of them moving as fast as $\sim 800-1000\,\kms$, a value that is consistent with the highest velocity material observed in LBGs at 
$2 \lta z \lta 3$ by \citet{Steidel10}.  The mass and metal-weighted outflow average speed typically ranges between 200 and 400 $\kms$. From 
redshift 9.3 to 3.0, the total 
mass of the main host halo increases from $5.0\times 10^{9}\,\msun$ to $2.4\times 10^{11}\,\msun$. The mean outflow speed increases for $5\lta z \lta 9.3$ and 
decreases again at $z\lta 5$, so there is no obvious correlation between halo mass and the peak outflow velocity in ErisMC. Similarly, we find only a weak correlation between the maximum wind velocity and star formation rate.  However, if we define the mass averaged mean outflow velocity at distance $r$ as: 

\begin{equation}
\langle{v_{\rm out}}\rangle(r) = \frac{\sum_{i}^{N}m_iv_{r,i}}{\sum_{i}^{N}m_i}
\end{equation}

where N is the total number of outflow gas particles in a radial shell of thickness $dr=$0.02 $R_{\rm vir}$ at distance $r$, $m_i$ is the mass of particle $i$ and $v_{r,i}$ its outflow radial velocity (relative to the host's center).  There is a correlation between the mean outflow velocity and the peak circular velocity of the host, as shown in Figure \ref{fig16}.
%This is consistent with results from Na I absorption-line surveys of galactic winds (cf. Figure 6 of \citealt{Veilleux05}) and follows 
%from the fact that, at increasing star formation rates, the higher SN energy input produces longer cooling shut-off times and larger outflows speeds. 

\subsection{Metallicity}

\begin{figure}
\centering
\includegraphics[width = 0.48\textwidth]{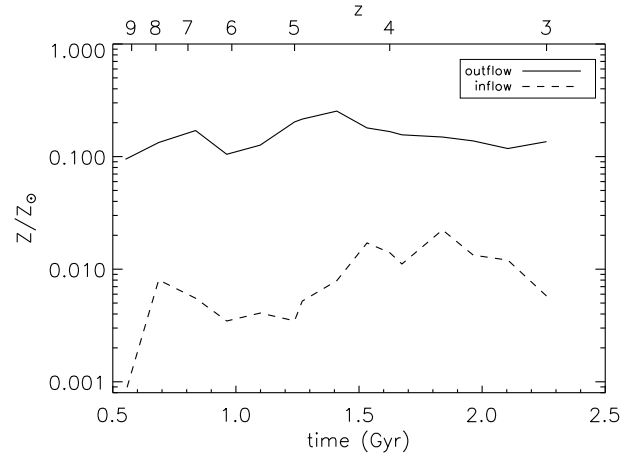}
\caption{Mean metallicity of inflowing ({\it dashed lines}) and outflowing ({\it solid lines}) material as a function of redshift. The average is taken over all
gas particles within a thin shell of radius $R_{\rm vir}$ and thickness $0.02R_{\rm vir}$.
}
\label{fig17}
\end{figure}

Figure \ref{fig17} shows the metallicity of inflowing and outflowing material at ErisMC's virial radius as a function of redshift. 
Galactic outflows are enriched to a typical metallicity in the range $0.1-0.2\,Z_\odot$ since redshift $\sim 9$, with little dependence on cosmic time. 
The mean metallicity of inflowing gas shows a more marked evolution, as it increases by about one dex in the interval $9\gta z\gta 3.5$. This is a consequence of more 
processed material falling back onto ErisMC in a ``halo fountain" \citep{Oppenheimer10} as well as being accreted via infalling satellites. Only half of the inflowing material 
at $R_{\rm vir}$ is unprocessed primordial gas. The gas metallicity of inflowing gas 
at z $\la 5$ ($Z\sim 0.01\,Z_{\odot}$) is typical of the metallicity observed in Damped Ly$\alpha$ and Lyman-Limit systems \citep[e.g.,][]{Wolfe05}. 

\subsection{Mass loading}

The mass loading factor characterizes the amount of material involved in a galactic outflow, and is defined as $\eta = \dot{M_{w}}/$SFR, where $\dot{M_{w}}$ 
is the rate at which mass is ejected. Observations of galactic outflows powered by starbursts suggest a wide range of mass loading factors, 
$\eta = 0.01 - 10$, with no obvious correlation with the star formation rates of their hosts \citep{Veilleux05}. In low-resolution cosmological simulations, the
mass loading factor is one of the input parameters \citep[e.g.,][]{Springel03,Oppenheimer06,DallaVecchia08,Choi11}. Most models adopt a constant mass loading 
factor, except for the momentum-driven model \citep{Oppenheimer06,Oppenheimer09}, where it is assumed to be inversely proportional to the host galaxy velocity
dispersion $\sigma$, $\eta=\sigma_{0}/\sigma$. In our SN-driven blastwave feedback scheme there is no specific parameter for mass loading. In this section we explore 
the mass loading factor in ErisMC and its variation with star formation and halo mass. 

\begin{figure}
\centering
\includegraphics[width = 0.48\textwidth]{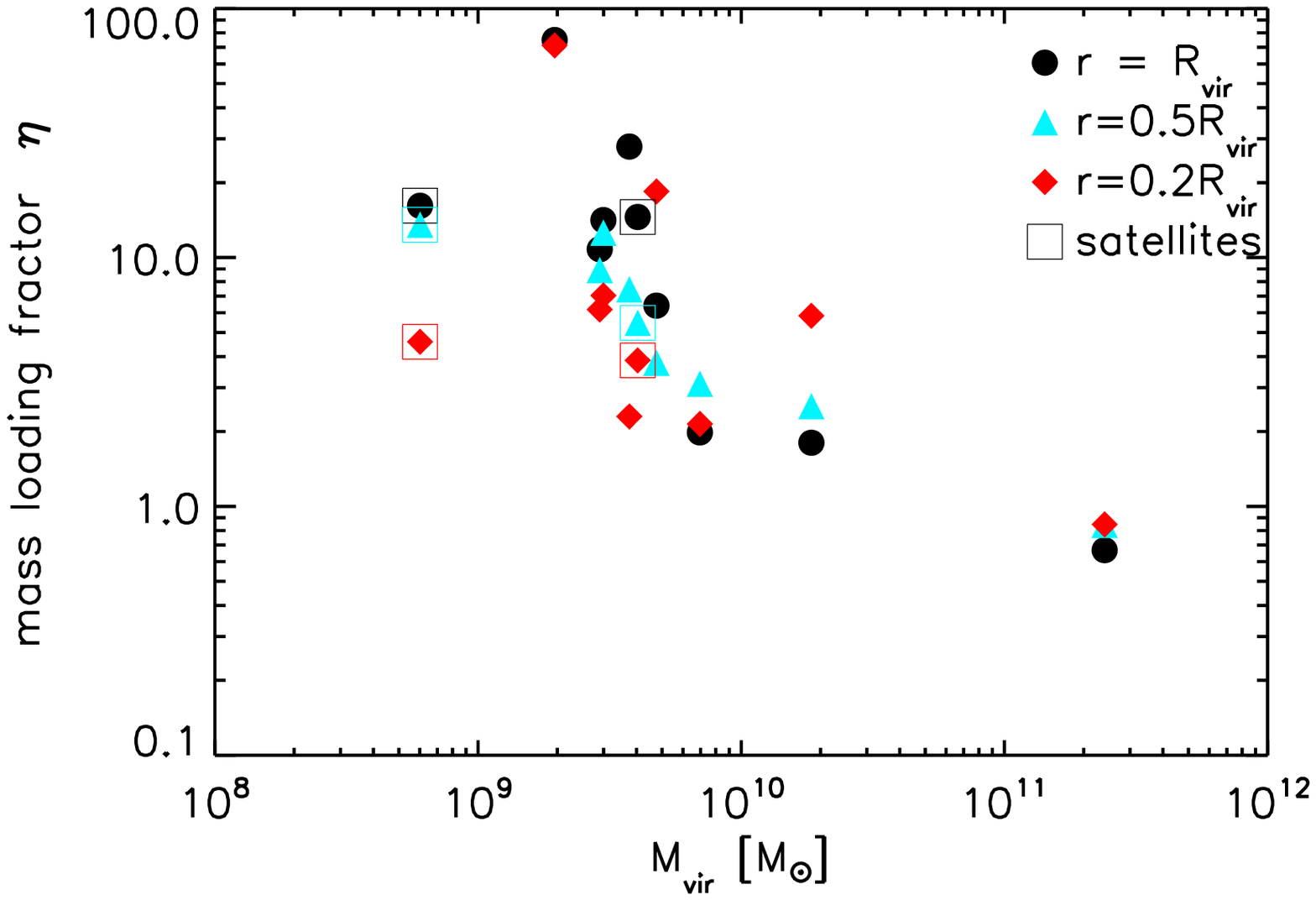}
\includegraphics[width = 0.48\textwidth]{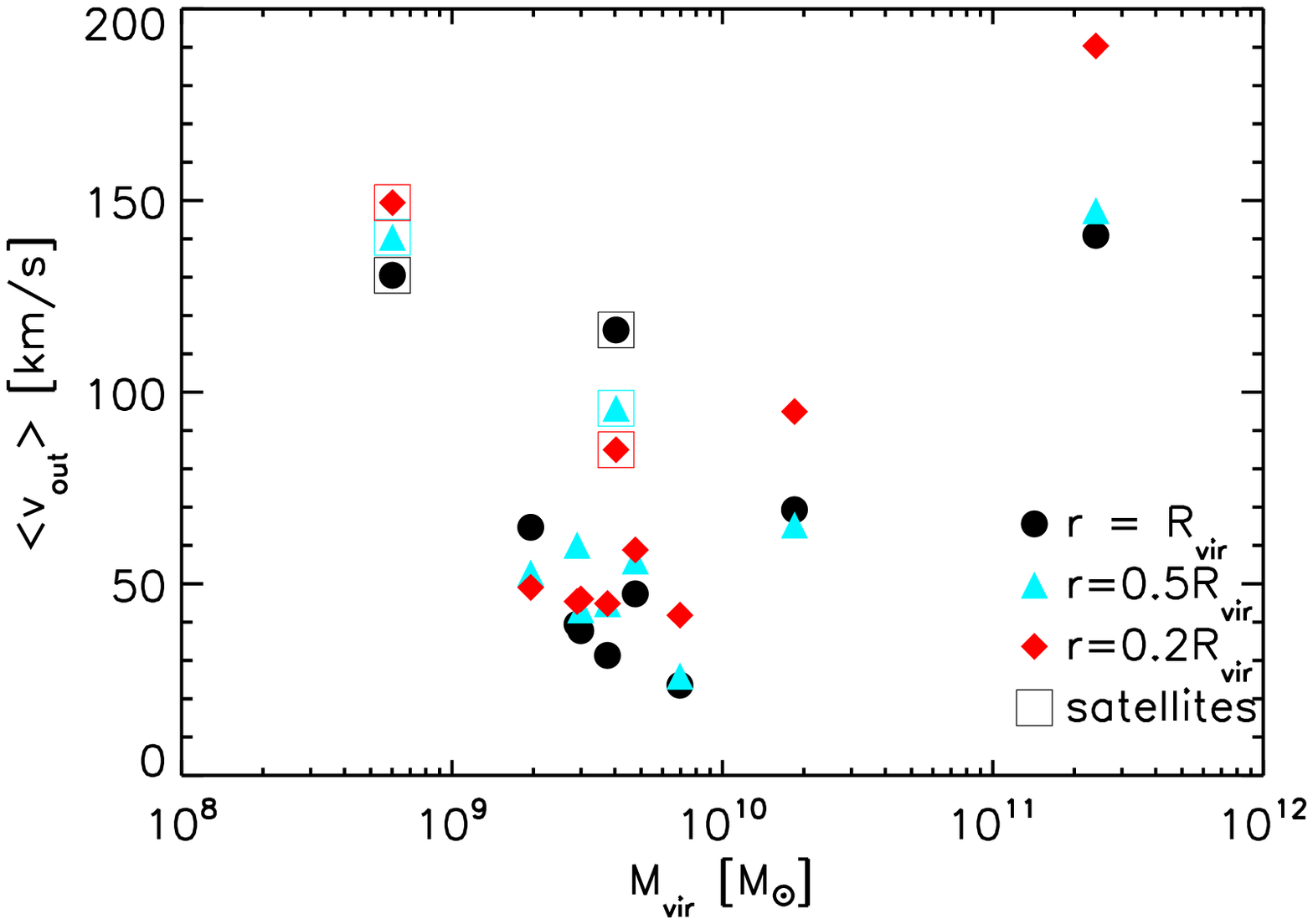}
\caption{{\it Top panel:} Mass loading factor at $z=3$ for the main host and the 9 most massive dwarfs within 250 kpc. The mass flux was measured at the virial 
radius ({\it black dots}, half the virial radius ({\it cyan triangles}) and one fifths of the virial radius ({\it red diamonds}). The empty square symbols
indicate satellite systems of the main host. {\it Bottom panel:} Same as the top panel, but for mass-weighted average outflow velocity $\langle v_{\rm out} \rangle$.   
}
\label{fig18}
\end{figure}

We compute the wind mass loading at a distance $r$ from the center as 
\begin{equation}
\dot{M_w}(r) = \frac{1}{\Delta r}\sum_{i}^{N}m_iv_{r,i}
\end{equation}
where $\Delta r$ is the thickness of the radial shell, N is the total number of outflow gas particles in the shell, $m_i$ is the mass of particle $i$ and $v_{r,i}$ its
outflow radial velocity (relative to the host's center). We have measured the mass flux at 0.2, half, and one virial radius, and
divided it by the instantaneous SFR ignoring any time delay between star formation and large-scale outflow, for ErisMC and the 9 most massive dwarf
halos within 250 kpc at $z=3$. We find (see Fig. \ref{fig18}) a strong correlation of the mass-loading factor ($\eta$) and the mean outflow velocity ($\langle v_{\rm out} \rangle$) with halo mass. While $\eta$ is of order unity for the main host, it can exceed 10 (and reaches 80 in one case) for nearby dwarfs and satellites. Similarly,  $\langle v_{\rm out} \rangle$ increases from $\sim$ 50 km/s for nearby dwarfs to $\sim$ 150 km/s for the main host.  Satellite systems of ErisMC are indicated with open square symbols in the figure, as their mass loss and outflow velocity may be affected by tidal stripping. It is interesting to note here that, with a comparable mass loading factor to ErisMC,  our Eris2 simulation \citep{Shen12} appears to be able to produce interstellar absorption line strengths of \Lya, \CII, \CIV, \SiII, and \SiIV as a function of galactocentric impact parameter that are in good agreement with those observed at high-redshift by \citet{Steidel10}.

\section{Summary}

We have presented a detailed study of the metal-enriched CGM of a massive galaxy at $z=3$ using results from a zoom-in hydrodynamic simulation 
of a disk galaxy with mass comparable to the Milky Way. Our approach to understanding the role of inflows, star formation feedback, and outflows in governing the gaseous 
and metal content of galaxies 
and their environment, is different to that of many recent theoretical efforts: Eris' extreme mass and spatial resolution allows us 
to follow self-consistently the venting of metals by small progenitor dwarf satellites and the transport of heavy elements from their 
production sites into the environment. The reference run adopts a blastwave scheme for supernova feedback that generates galactic outflows 
without explicit wind particles, and a star formation recipe based on a high gas density threshold.
%and a model for the diffusion of metals and thermal energy.

We have found that ErisMC's metal-enriched CGM extends as far as 4 virial radii (about 200 physical kpc) from its center.
%The volume filling factor of metal-enriched material within 200 kpc from the center is about 10\%. 
Approximately 41\%, 9\%, and 50\% of all gas-phase heavy elements within 250 kpc from the center are hot ($T>3\times 10^5$ K), warm ($3\times 10^5 {\rm K}>T>3\times 10^4$ K), 
and cold ($T<3\times 10^4$ K), respectively. More than 40\% of all gas-phase metals lie outside the virial radius: while cold metal-rich material 
traces large overdensities within the main host, about 50\% of all warm and 70\% of all hot metals are found in 
low density $\delta < 30$ regions beyond ErisMC's virial radius. 
Intergalactic metals are characterized by a strong temperature gradient with overdensity,
as the metal-weighted temperature climbs from $10^4$ K at $\delta=1$ to above $2\times 10^5$ K at $\delta=10$.     
SN-driven winds are able to transport metals to regions at $\delta\lta 10$ where they accumulate, creating a peak in the cold 
gas metallicity-density relation at $Z/Z_\odot\approx 0.05$. The mean metallicity of cold gas drops quickly below $\log Z/Z_\odot=-2$ 
at overdensities $\delta\lta 10$.

We have identified three sources of heavy elements in the simulated region: 1) the main host,
responsible for 60\% percent of all the metals found within $3\rvir$, and for none of those found beyond $3\rvir$; 2) its satellite progenitors
-- systems accreted by the main halo before redshift 3, which shed their metals before and during infall and are responsible for 28\% of 
all the metals within $3\rvir$; and its orbiting nearby dwarfs, which give 
origin to 12\% of all the metals within $3\rvir$ and 95\% of those beyond $3\rvir$. Late ($z<5$) galactic ``superwinds" -- the result 
of recent star formation in ErisMC -- account for only 9\% of all the metals observed beyond $2\rvir$, the rest having been 
released at redshifts $5\lta z \lta 8$. These findings confirm the ideas put forward by \citet{Porciani05}, that enrichment from nearby dwarfs 
may contribute significantly to (and dominate at large distances) the pollution of the CGM around LBGs.    
\citet{Porciani05} argued that massive galaxies at high redshift are likely to be surrounded by the metal bubbles produced by nearby dwarfs and satellites
at earlier epochs, and that gravity will tend to increase the spatial association between such metal bubbles and LBGs.
The strong association observed between stronger \CIV\ systems and LBGs led \citet{Adelberger03} to argue that metal-rich ``superwinds" 
from LBGs may be responsible for distributing the product of stellar nucleosynthesis on (comoving) Mpc scales. The analytical results of 
\citet{Porciani05} and the simulations presented here show that this is not the case, as it is nearby dwarfs that dominate the metal enrichment beyond 2-3 $R_{\rm 
vir}$ of LBGs. This is not to imply that most of the metals observed in the IGM at high redshift have been ejected by 
dwarf galaxies: rather than, as a function of distance from a massive system, the contribution of the main host becomes sub-dominant compared to that of 
its smaller companions. Simulations of larger cosmological volumes are needed to assess whether our results are consistent with observations 
of the diffuse metal-enriched IGM.

Substantial amounts of heavy elements are generated at a larger distance from the main host's center than their current location, and subsequently are 
accreted by the host along filaments via low-metallicity cold inflows. Galactic outflows have velocities of a few hundred $\kms$. The outflows 
decelerate rapidly, and the resulting long metal transport timescales produce an age gradient in metals as a function of distance from the
main host. The outflow mass-loading factor is of order unity in the main halo, but can exceed a value of 10 for nearby dwarfs. 

As a Lagrangian particle method, SPH does not include any implicit diffusion of scalar quantities such as metals. In the absence of some 
implementation of diffusion, metals are locked into specific particles and their distribution may be artificially 
inhomogeneous \citep{Aguirre05,Wiersma09}. A number of simulations of even higher resolution than ErisMC and including a scheme 
for turbulent mixing that redistributes heavy elements and thermal energy between the outflowing material and the ambient gas \citep{Shen12}
are in the making. 

\acknowledgments
Support for this work was provided by the NSF through grant AST-0908910 and OIA-1124453 (P.M.). Simulations were carried out on NASA's Pleiades supercomputer and 
the UCSC Pleiades cluster. One of us (P.M.) acknowledges support from a Raymond and Beverly Sackler Visiting Fellowship at the Institute of Astronomy, Cambridge.

%\clearpage

\end{document}